\documentclass[aps,pre,twocolumn,showpacs,superscriptaddress]{revtex4-1}

\usepackage{amsmath}
\usepackage{amssymb}
\usepackage{bbold} 
\usepackage{mathtools}
\usepackage{graphicx}
\usepackage{hyperref}
\usepackage[arrow, matrix, curve]{xy}
\usepackage{bm}
\usepackage{bbm} 
\usepackage{yhmath}
\usepackage{color}
\usepackage{url}
\newcommand\diff{\mathrm{d}}

\usepackage[usenames,dvipsnames]{xcolor}
\hypersetup{colorlinks=true, linkcolor=BrickRed, urlcolor=blue!50!black, citecolor=blue!50!black}

\usepackage[normalem]{ulem}

\makeatletter
\newcommand\hide@visible[1]{%
  \bgroup\fboxsep=.3ex\colorbox{Gray}{begin hide}%
  #1\colorbox{Gray}{end hide}\egroup%
}
\newcommand\hide@hidden[1]{%
  \bgroup\fboxsep=.3ex\colorbox{Gray}{hidden text}%
}
\newcommand\hide@invisible[1]{}
\newcommand\makevisible{\let\hide\hide@visible}
\newcommand\makehidden{\let\hide\hide@hidden}
\newcommand\makeinvisible{\let\hide\hide@invisible}
\makeatother
\makehidden

\usepackage[capitalise,nameinlink]{cleveref}

\crefname{section}{Sec.}{Secs.} 
\usepackage{graphicx}
\usepackage[caption=false]{subfig} 
\usepackage{color}
\usepackage{verbatim} 
\usepackage{natbib}

\begin{document}

\title{Intermediate scattering function of colloids in a periodic laser field}
\altaffiliation{This publication is in honour and remembrance of our friend, mentor, and colleague Stefan U. Egelhaaf. }
\author{Regina Rusch}
\altaffiliation{R.R and Y.S. contributed equally to this work. }
\affiliation{Institut f\"ur Theoretische Physik,  Universit\"at Innsbruck, Technikerstra{\ss}e 21-A, 6020 Innsbruck, Austria.}
\author{Yasamin Mohebi Satalsari}
\altaffiliation{R.R and Y.S. contributed equally to this work.}
\affiliation{ Condensed Matter Physics Laboratory, Heinrich Heine University, Universit\"atsstraße~1, 40225 D\"usseldorf, Germany.}
\author{Angel B. Zuccolotto-Bernez}
\affiliation{ Condensed Matter Physics Laboratory, Heinrich Heine University, Universit\"atsstraße~1, 40225 D\"usseldorf, Germany.}
\author{Manuel A. Escobedo-Sánchez}
\email{E-mail: escobedo@hhu.de}
\affiliation{ Condensed Matter Physics Laboratory, Heinrich Heine University, Universit\"atsstraße~1, 40225 D\"usseldorf, Germany.}
\author{Thomas Franosch}
\email{E-mail: thomas.franosch@uibk.ac.at }
\affiliation{Institut f\"ur Theoretische Physik,  Universit\"at Innsbruck, Technikerstra{\ss}e 21-A, 6020 Innsbruck, Austria.}

\date{\today}

\begin{abstract}
We investigate the dynamics of individual colloidal particles in a one-dimensional periodic potential using the intermediate scattering function (ISF) as a key observable.
We elaborate a theoretical framework and derive formally exact analytical expressions for the ISF. We introduce and analyze a generalized ISF with two wave vectors to capture correlations in a periodic potential beyond the standard ISF. Relying on Bloch's theorem for periodic systems and, by solving the Smoluchowski equation for an overdamped Brownian particle in a cosine potential, we evaluate the ISF by numerically solving for the eigenfunctions and eigenvalues of the expression. We apply time-dependent perturbation theory to expand the ISF and extract low-order moments, including the mean-square displacement, the time-dependent diffusivity, and the non-Gaussian parameter. Our analytical results are validated through Brownian-dynamics simulations and experiments on 2D colloidal systems exposed to a light-induced periodic potential generated by two interacting laser beams.
\end{abstract}

\maketitle

\section{Introduction}
Colloidal particles suspended in fluids serve as a powerful model system in soft matter and biological physics. Their microscopic size makes them easy to manipulate, allowing researchers to explore fundamental physical principles in a controlled environment\cite{Poon_1996,Potiguar_2007,Bechinger2014ColloidsOP,Perez_2021,Xiaoguang_2015, Egelhaaf_2011, Castaneda_2021}. A major breakthrough in this field was the invention of optical tweezers by Ashkin et al., which enabled the precise trapping and manipulation of individual particles using focused laser beams \cite{ashkin1970acceleration, ashkin1980applications,ashkin1997optical}. 
Building on this advancement, scientists have since developed laser interference patterns to create structured external potentials, offering a unique way to study how particles behave under spatially periodic forces. These optical potentials arise from the interaction between light and particles with a refractive index different from that of the surrounding medium, creating highly tunable energy landscapes\cite{Alexandr_2009,gao2017optical}.

In order to understand how colloidal particles move within these periodic potentials, researchers have analyzed the probability distribution, the first-passage time,  and low-order moments, such as the mean-square displacement, the long-time diffusivity, and the non-Gaussian parameter \cite{Egelhaaf_2011,Perez_2021,jenkins2008colloidal,capellmann2018dense}. Although these observables provide valuable information, most theoretical studies have focused only on diffusion coefficients in the short- and long-time limits \cite{Lifson_1962, Festa_1978, Stratonovich_1967_topics}. More recent work has expanded on these studies by analytically exploring tilted washboard potentials \cite{Reimann_2002,
Reimann_2001}, comparing theoretical predictions with experimental results \cite{Reimann_2008}, and investigating memory effects in such systems \cite{Fulde_1975,Straube_2024}. These low-order moments are useful for identifying deviations from free Brownian motion, revealing effects such as trapping and non-Gaussianity\cite{Kurzthaler_2016, Xu_2024}. However, they offer only a limited perspective of the system dynamics, as they do not capture the full spatial-temporal evolution of particle motion. To achieve a more complete description, it is necessary to adopt a framework that incorporates both spatial and temporal correlations.

One such framework emerges naturally in Markovian systems, where the future state depends only on the present and not on the past. In these systems, all relevant dynamical information is contained in the propagator, which describes the probability of a particle transitioning between states over a given time interval. Although the propagator provides a full statistical description of particle motion, it is often challenging to access experimentally. A more practical and experimentally accessible alternative is the intermediate scattering function (ISF), which encodes both spatial and temporal correlations in particle motion. Unlike traditional low-order moments, the ISF offers a more comprehensive characterization of the system and can be directly measured using techniques such as dynamic light scattering\cite{berne2013dynamic,dhont1996introduction}, differential dynamic microscopy \cite{cerbino2008differential}, and single-particle tracking \cite{CrockerGrier1996}. 

Furthermore, upon a small wave vector expansion of the ISF the low-order moments are recovered. Analytical derivations of the ISF have been achieved for various systems, including anisotropic active Brownian particles\cite{Kurzthaler_2016, Kurzthaler_2018}, Brownian circle swimmers in gravitational fields \cite{Kurzthaler_2017,Chepizhko_2022,Rusch_2024}, anisotropically diffusing colloidal dimers\cite{Mayer_2021}, and run-and-tumble agents \cite{Angelani_2013,Kurzthaler_2024}. Notably, experimental validations have been conducted for colloidal dimers and active colloids.

In this work, we extend the traditional ISF by introducing a generalized version that incorporates two wave vectors, allowing us to investigate the correlations in periodic systems in reciprocal space. The central question we address is twofold. First, we develop a theoretical framework to gain analytic insight into the dynamics of colloidal particles in general periodic potentials. Second, we compare our predictions with experimental results, exploring a spatio-temporal regime that has not yet been fully investigated. 

To develop the theoretical framework, we solve the Smoluchowski equation for a single, overdamped Brownian particle in a cosine potential. Reformulating the problem as a Hermitian Schr\"odinger equation allows expressing the solutions in terms of eigenvalues and eigenfunctions \cite{risken1996fokker}. Taking advantage of the system’s periodicity, we apply Bloch’s theorem, which provides a systematic way to expand the ISF using time-dependent perturbation theory and extract key dynamical moments. These methods are based on previous work  \cite{Chepizhko_2022,Rusch_2024,Lapolla_2020} and have been successfully applied to bistable periodic potentials \cite{Asaklil_1999,pattanayak2024impact}.  To validate our theoretical predictions, we compared our theoretical framework with Brownian dynamics simulations.

Experimentally, We track individual colloidal particles confined between two walls in a two-dimensional configuration, subjected to a periodic potential created by the interference of two laser beams. By systematically varying the laser power, we control the potential amplitude and record particle trajectories under different conditions. The generalized ISF, along with its associated moments, is then extracted from these experimental data and directly compared with our theoretical predictions, enabling a precise evaluation of the Brownian motion description in periodic optical fields.  
By integrating theoretical modeling, experimental measurements, and computational simulations, this work aims to provide a comprehensive understanding of colloidal motion in structured environments. Our approach bridges the gap between fundamental stochastic dynamics and experimentally accessible observables, offering new insights into the interplay between Brownian motion and periodic potentials.

This work is organized as follows. 
In \cref{sec:experimental_setup} the experimental materials and methods are described. In \cref{sec:theory}, we introduce the analytical model, and derive the theoretical framework and observables. Readers more interested in the results rather than the theoretical derivation may skip this section and jump directly to \cref{sec:Results}, where we present our findings, compare experimental results with theoretical predictions, and discuss their implications. Finally, we summarize our findings, concluding with an outlook on future research directions in this field in \cref{sec:conclusion}.

\section{Material and methods} \label{sec:experimental_setup}

This section describes the employed numerical methods and the used experimental setup to study colloidal dynamics in periodic potentials. We describe the processes of preparing two-dimensional colloidal samples and explain our custom-built optical setup that generates periodic laser fields. Also, the experimental protocol, including data acquisition parameters and particle tracking methodology, is described. Our experimental design allows us to systematically investigate the behavior of colloidal particles under varying potential strengths while maintaining high spatial and temporal resolution, which is essential for a rigorous comparison with the theoretical framework presented in \cref{sec:theory}.
\subsection{Numerical methods}
To complement our analytical and experimental approach, we conducted Brownian-dynamics simulations of single particles in a one-dimensional cosine potential. We implemented the Euler-Maruyama method \cite{maruyama1955continuous} for numerically solving the differential equation of motion. For efficient sampling of observables, we used the extended order-n algorithm of Frenkel and Smit, which enables equidistant sampling on a logarithmic time scale \cite{frenkel2002understanding,siems2017computersimulationen,Siems_2018}.

To numerically solve the equation of motion, as it will be explained in detail in the theory \cref{sec:theory}, we employed a spectral method. Specifically, we expanded the equation in its Fourier basis to obtain its operator form. Numerically, we truncated the expansions to form a finite-dimensional matrix eigenvalue problem. We used the scipy.linalg.eig function from the SciPy library \cite{virtanen2020scipy} to compute the eigenvalues and eigenvectors of the resulting matrix. The convergence of our results was verified by systematically increasing the truncation order until the eigenvalues stabilized to the desired precision. 

\subsection{Sample preparation}\label{SamplePreparation}

The experiments were performed with two-dimensional (2D) colloidal systems to ensure controlled particle motion mainly in the horizontal plane. We prepared a dilute suspension of monodisperse polystyrene sulfate latex particles (radius 1.5 \textmu m, 4$\%$ polydispersity, Thermo Fisher Scientific, batch number 1660463) in ultra-pure water (resistivity 18.2 M$\Omega$cm, Purelabs Flex, Elga).  To build the sample cells chamber, we first attached a rectangular coverslip (Thickness No. 1, 24 × 50 mm, VWR 631-0146) to a microscope slide using UV-curing adhesive (NOA61, Norland Products Inc.). We then carefully deposited 2.5 $\mu$L of the particle suspension at the center of this base coverslip. A smaller square coverslip (Thickness No. 1.5, 22 × 22 mm, VWR 631-0125) was gently placed on top to create a thin chamber. We sealed the edges with UV-curing adhesive to prevent evaporation during long observation periods. To ensure enough space between the coverslips, particles with a radius of 2 \textmu m were added to the colloidal suspension to act as spacers. These spacers ensure a sufficient gap between coverslips that counteracts the pull of capillary forces, preventing the coverslips from coming close to each other and squeezing the main particles. The resulting configuration reached an area fraction of $0.01$. The completed cell was positioned on the microscope stage for observation.

\subsection{Experimental Setup}\label{Experimental Setup}
\begin{figure}[!ht]
    \centering
    \includegraphics[width=\linewidth]{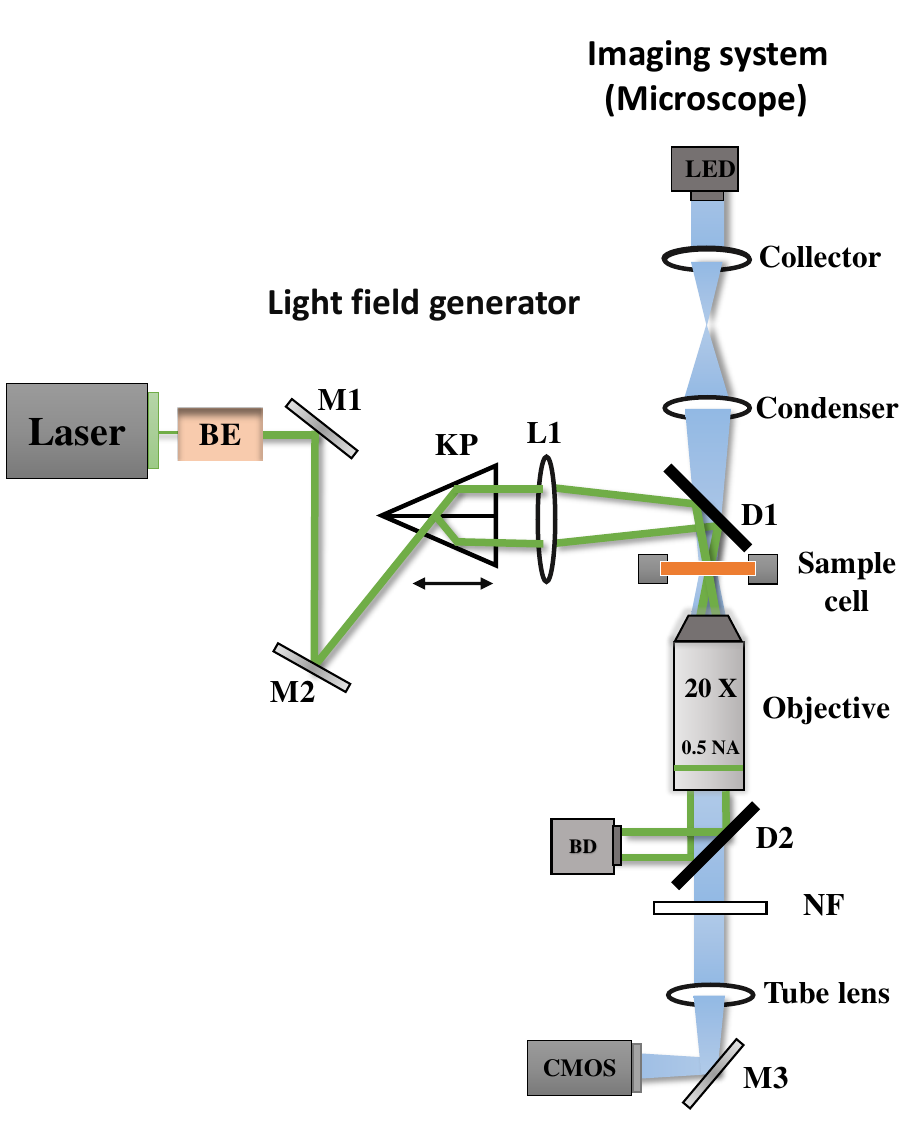}
    \caption { Illustration of the experimental setup. The laser field generator utilizes a 532 nm laser beam to create a periodic light field through a K\"oster prism (KP). The two parallel beams from KP are focused on the sample plane using lens $L1$ and their interference creates the fringe pattern. The imaging system, consisting of a microscope with a 20× objective and a CMOS camera, captures the sample's particle trajectories.}
    \label{fig:Fig1}
\end{figure}

Our experimental setup consists of a laser field generator and an imaging system, as illustrated in \cref{fig:Fig1}.
The laser field generator creates the periodic potential experienced by the colloidal particles. A 532 nm laser (Cobolt 05-01 Samba, 1.5 W) beam passes through a beam expander and is directed by mirrors (M1 and M2) to a K\"oster prism (KP). This prism splits the incoming beam into two parallel beams of equal intensity, which are then focused by lens L1 and directed by a dichroic mirror (D1) into the sample plane. At the sample plane, the beams interfere to generate the laser field inducing a periodic potential, with a periodicity of L=($4.15\pm0.07$) \textmu m.
By adjusting the position of the K\"oster prism, we can fine-tune this periodicity. Additionally, by controlling the laser power, we can precisely modulate the amplitude of the potential experienced by the particles. The relationship between the amplitude of the potential as a function of laser power (calibration) is presented in \cref{app:Calibration}.

The imaging system is an inverted bright-field microscope (Nikon Ti-E) with a 20× objective (Nikon Plan Flour, 0.5 NA). We illuminate the sample with a blue LED (Thorlabs M455L4) and capture images using a CMOS camera (Mako U-130B) at a resolution of 1280×1024 pixels, with a pixel pitch of 0.24 \textmu m/px. To prevent laser light from reaching the camera sensor, we employed a second dichroic mirror (D2) that redirects the laser beams to a beam dump (BD), with any residual laser light being filtered out by a notch filter (NF). This optical arrangement allows us to simultaneously apply the periodic potential and observe particle dynamics with high precision.

\subsection{Experimental realization}\label{Experimental relization}
We conducted all experiments in a temperature-controlled laboratory environment maintained at 20°C. Before measurements, samples were allowed to equilibrate for 48 hours to ensure thermal stability. For each experiment, we recorded 180,000 images over 2 hours, using a frame rate of 25 Hz and an exposure time of 1 ms. These parameters represent a balance between temporal resolution and the need to collect sufficient data for robust statistical analysis. We collected between 400 and 1500 individual particle trajectories for each laser power from several measurements. At the highest laser power, the pressure radiation pushed approximately 20$\% $ of particles out of the field of view, requiring us to sample multiple regions within each sample cell to obtain statistically meaningful datasets.
To extract particle positions with high precision, we employed a custom MATLAB-based tracking algorithms adapted from the work of D. Blair and E. Dufresne \cite{Matlabroutine}  together with the Michalet algorithm \cite{Michalet2010}, achieving a localization uncertainty of approximately 3 nm. Trajectories are openly available in Zenodo at https://doi.org/10.5281/zenodo.14931759.

This exceptional spatial resolution is crucial for accurately capturing the subtle dynamics of particles within the periodic potential, particularly when comparing experimental results with theoretical predictions. The observables are computed for each measurement and averaged over all the measurements. Results in \cref{sec:Results}, represent the mean values and the error bars their standard deviation.

\section{Theory and observables} \label{sec:theory}
In this section, we introduce the model and the theoretical framework. Further, we derive the generalized ISF and low-order moments. From the expansion of the ISF, we compute exact expressions of the mean-square displacement, diffusivity, and non-Gaussian parameter. 

\subsection{Model}\label{sec:Model}
We model the interaction of the colloid and the interfering laser beam by a one-dimensional periodic potential $U(x) = U(x+L) $ with period $L$. We assume a simple cosine potential
\begin{align} \label{eq:cosine}
    U(x)= U_1 \cos (Q_1 x),
\end{align} 
where $Q_1=2\pi/L$ is the wave vector associated with the period $L$. We introduce the dimensionless amplitude $u=U_1/k_BT$ where $k_B T$ is the thermal energy scale. 
The colloid is assumed to undergo Brownian motion in the presence of the spatially periodic force $-\partial_x U(x) $. Then the equation of motion for its position $x(t)$ is provided by the Langevin equation
\begin{align} \label{eq:Langevin}
	\dot{x}(t) 
	&=  D Q_1  u \sin(Q_1 x(t))+ \eta(t).
\end{align}
Here, $D$ denotes the short-time diffusion coefficient such that $D/k_B T$ is the mobility as derived by the Einstein relation. The stochastic term $\eta (t)$ represents a  centered Gaussian white noise with a delta-correlated variance $\langle \eta (t) \eta (t^\prime) \rangle = 2 D \delta (t-t^\prime)$. 

We identify three quantities which set the characteristic units of the system: The period $L$ of the potential is the fundamental unit of length. The time a free particle needs to diffuse the distance of a period is $L^2/D$ and will be used as a time unit. Energies are compared to the thermal energy $k_B T$. Therefore, the problem displays the dimensionless amplitude $u$ as a single control parameter. 

\subsection{Smoluchowski equation}

To make analytic progress, we rely on the Smoluchowski equation which is equivalent to the Langevin description in Eq.~\eqref{eq:Langevin} in  the case of 
equilibrium dynamics. Here, the fundamental quantity is the propagator $\mathbb{P} \coloneq \mathbb{P}(x,t|x_0)$, defined as the conditional probability of finding the particle at position $ x $ at time $ t $, given its initial position was $ x_0 $ at time zero. The initial condition is therefore provided by $\mathbb{P}(x,t=0|x_0)= \delta(x-x_0)$. By standard methods~\cite{risken1996fokker} one derives the Smoluchowski equation governing the time evolution of the propagator
\begin{align} \label{eq:FP}
	\partial_t \mathbb{P} 
	= -\partial_x \left[  D  Q_1 u \sin(Q_1 x )\mathbb{P} \right] + D \partial_x^2 \mathbb{P} 
	\eqcolon \Omega \mathbb{P}.
\end{align}
A stationary solution is provided by 
\begin{align} \label{eq:stationary_solution}
	p^{\text{eq}}(x) =\frac{1}{Z_1} \exp[- U(x) /k_B T] ,
\end{align}
with a normalization factor $Z_1$. Since in the infinite system no stationary solution exists, we choose $Z_1$ as the normalizing factor for a unit cell
\begin{align}\label{eq:normalize_peq}
\int_0^L \diff x \, p^{\text{eq}}(x) = 1.
\end{align}
Explicitly
\begin{align}
	Z_1  =  \int_0^L\! \diff x\, e^{- u \cos(Q_1 x) } 
	= L\,  \mathrm{I}_0( u),
\end{align}
where $\mathrm{I}_\nu(\cdot)$ denotes the modified Bessel function of the first kind to order $\nu$. 

We replace our infinite system with a finite system by dividing it into $N \in \mathbb{N}$ unit cells of length $L$ and apply periodic boundary conditions.  The limit $N \to \infty$ will be performed at the end of the calculations. 

In order to find  non-trivial solutions of \cref{eq:FP} we first perform a separation ansatz
\begin{align}
	\mathbb{P} &=  E(t)\psi(x),
\end{align}
and find the solution for the time-dependent part immediately as $E(t) = e^{-\lambda t} $. For the position-dependent part, we have to solve an eigenvalue equation.
As the Smoluchowski operator $\Omega $ is  non-Hermitian, we distinguish between \textit{right} and \textit{left} eigenfunctions
\begin{align}  \label{eq:eigenvalue_FP_R}
 	\Omega \psi^R_\lambda(x) &= - \lambda \psi^R_\lambda(x)  , \nonumber \\
	\Omega^\dagger \psi^L_\lambda(x) &= - \lambda^* \psi^L_\lambda(x)  ,
\end{align} 
and $\Omega^\dagger$ is the adjoint operator with respect to the scalar product 
\begin{align} \label{eq:scalar_product}
	\langle \phi |\psi \rangle := \frac{1}{N}  \int_0^{NL} \phi(x)^* \psi(x)   \diff x .
\end{align}

Left and right eigenfunctions to different eigenvalues are orthonormal, satisfying
\begin{align} \label{eq:orthonormality_FP}
	\langle \psi_\lambda |\psi_{\lambda'} \rangle = \frac{1}{N} \int_0^{NL} \psi_\lambda^L(x)^* \psi_{\lambda'}^R(x)   \diff x = \delta_{\lambda\lambda'} .
	\end{align}
Therefore, only the product of left and right eigenstates to identical eigenvalue is normalized. Furthermore, the eigenfunctions are complete, fulfilling the condition
	\begin{align} \label{eq:completeness_FP}
		\frac{1}{N} \sum_{\lambda}^\infty \psi_\lambda^R(x)   \psi_\lambda^L(x_0)^*   = \delta(x-x_0) .
	\end{align} 
By comparing \cref{eq:FP,eq:eigenvalue_FP_R} for the stationary state, we identify that the eigenfunction $\psi^R_0(x)$ to eigenvalue zero  has to be proportional to the equilibrium distribution $\psi_{0}^R(x) \propto p^\text{eq}(x) $. 
We choose 
\begin{align} \label{eq:stationary_SE}
	\psi_0^R(x) = p^\text{eq}(x), \qquad  \psi_0^L(x)^*=1,
\end{align}
which fulfills the  normalization conditions \cref{eq:orthonormality_FP,eq:normalize_peq}.

The formal solution of the Smoluchowski, \cref{eq:FP}, can be written as $\mathbb{P}=e^{\Omega t}\delta(x-x_0)$. We can insert the completeness relation, \cref{eq:completeness_FP}, and apply the eigenvalue equation, \cref{eq:eigenvalue_FP_R}, and obtain 
\begin{align} \label{eq:probability_density_FP}
\mathbb{P}(x, t| x_0)  = \frac{1}{N}  \sum_{\lambda}^\infty e^{- \lambda t} \psi_\lambda^R(x)  \psi_\lambda^L(x_0)^*.
\end{align}

\subsection{Schr\"odinger form}

It is favorable to transform the Smoluchowski equation, \cref{eq:FP}, into a Schr\"odinger-like equation using the `gauge transform'~\cite{risken1996fokker}
\begin{align}
	\mathbb{P} &=  e^{- U(x)/2 k_B T}\mathbb{P}_0  .
\end{align}
 A straightforward calculation reveals that  this yields in general to
\begin{align}
	\partial_t \mathbb{P} _0=&   D \partial^2_x \mathbb{P}_0 - D\frac{[U^\prime(x)]^2}{4 (k_B T)^2}  \mathbb{P}_0 
	+ \frac{DU^{\prime\prime}(x)}{2 k_B T} \mathbb{P}_0.
\end{align}
By specializing to the potential given in \cref{eq:cosine}, we find
\begin{align} \label{eq:FPschroedinger}
	  \partial_t \mathbb{P}_0  	&= \frac{D}{L^2} \left[ 2\pi^2 u \cos( Q_1 x) - \pi^2 u^2  \sin^2(Q_1 x)  \right] \mathbb{P}_0  +D   \partial^2_x \mathbb{P}_0 \nonumber \\
	  &
	\eqcolon \mathcal{L}_0 \mathbb{P}_0.
\end{align}
We note that the operator $\mathcal{L}_0$ is \textit{Hermitian} with respect to the standard scalar product in \cref{eq:scalar_product}.
Equivalently to the procedure of finding solutions of the Smoluchowski operator, to find the non-trivial solution of \cref{eq:FPschroedinger} we perform again a separation ansatz
\begin{align}
	\mathbb{P}_0 &=  E(t)\Psi(x),
\end{align}
with $E(t) = e^{-\lambda t} $ and for the position-dependent part, we have to solve an eigenvalue equation
\begin{align}
	\mathcal{L}_0 \Psi_\lambda(x) = - \lambda \Psi_\lambda(x)  ,
\end{align}  
where $ \lambda $ represents the eigenvalue and $\Psi_\lambda(x)$ the eigenfunction.  
We note that the transformation of the eigenfunctions of the Schr\"odinger operator to the Smoluchowski operator \cref{eq:eigenvalue_FP_R} is provided by
\begin{align} \label{eq:eigenfunc_transform}
\psi^R_{\lambda}(x)=\Psi_{\lambda}(x)\sqrt{p^\text{eq}(x)} , \quad
\psi^L_{\lambda}(x)=\Psi_{\lambda}(x)/\sqrt{p^\text{eq}(x)}  .
\end{align} 
As $\mathcal{L}_0 $ is an Hermitian operator, the eigenvalues are real and two eigenfunctions with different eigenvalue are orthonormal, equivalently to \cref{eq:orthonormality_FP}.  
The eigenfunctions are complete, fulfilling the completeness relation, similarly to \cref{eq:completeness_FP}.
The eigenfunction to eigenvalue zero can be easily found using 
\cref{eq:stationary_SE,eq:eigenfunc_transform} 
\begin{align} \label{eq:eigenfunctions_0_S}
	\Psi_{0}(x) = \Psi_{0}(x)^{*} = \sqrt{p^\text{eq}(x)}. 
\end{align}
\subsection{Bloch representation}
For periodic systems we make use of the Bloch representation of the eigenfunctions in terms of the discrete wave vector $q \in (2\pi/NL ) \mathbb{Z}$, with $-\pi/L < q \leq \pi/L$ and the discrete band index $n$. The wave vector can be restricted to the first Brillouin zone (BZ) because of the periodicity of the systems. The  eigenfunctions are of the form
	\begin{align}
		\Psi_{nq}(x) = e^{i q x} u_{nq}(x)  ,
	\end{align}
	where the Bloch function $u_{nq}(x) = u_{nq}(x+L)$ is periodic, and the wave functions obtain a second index $q$, representing the wave vector. The associated eigenvalue will be denoted by $\lambda_{nq}$. The orthonormality relation for the Bloch functions can be expressed as
		\begin{align}\label{eq:ortho}
			\int_0^{L} \! \diff x \, u_{nq} (x)^{*} u_{mq} (x) = \delta_{nm}, 
		\end{align}
		see \cref{app:orthonormality_simplification} for the detailed derivation. Also, the completeness relation still holds 
	\begin{align} \label{eq:completeness_u}
		\delta(x-x_0)  &= \sum_{n} u_{nq}(x) u_{nq}(x_0) ^{*},
	\end{align}	
where $x$ and $x_0$ are within the same unit cell.
Finally, we obtain the probability density using \cref{eq:probability_density_FP} and transforming the eigenfunction of the Smoluchowski operator to the eigenfunctions of the Schr\"odinger operator with \cref{eq:eigenfunc_transform}. Lastly, inserting the Bloch form of the eigenfunctions we find the probability density
	\begin{align} \label{eq:probability_density_S}
	 &\mathbb{P}(x, t| x_0)  = \frac{1}{N}\sqrt{\frac{p^\text{eq}(x)}{p^\text{eq}(x_0)}} \sum_{q \in \text{BZ}} \sum_{n}  e^{-\lambda_{nq}t}\Psi_{nq}(x) \Psi_{nq}(x_0)^{*}  \nonumber \\
	 &=  \sqrt{\frac{p^\text{eq}(x)}{p^\text{eq}(x_0)}}\int_{\text{BZ}} \!\frac{ L \, dq }{2\pi}  \sum_{n}  e^{-\lambda_{nq}t} e^{i q (x-x_0)}  u_{nq}(x) u_{nq}(x_0)^{*},
 \end{align}
 where in the last step the thermodynamic limit $N\to\infty$ was performed and therefore, the sum over the wave vectors can be replaced by  an integral over the Brillouin zone (BZ).

\subsection{Dirac notation}
To make further progress, we introduce the compact Dirac notation where we rely on the isomorphism between the periodic square-integrable functions $u_{nq}(x)  \in L^2[0,L]$ and the abstract kets $|u_{nq}\rangle $ in a separable Hilbert space  $\mathcal{H}$. We introduce the generalized position basis $|x\rangle$, such that $ u_{nq}(x) = \langle x | u_{nq} \rangle$. In particular, the set of associated eigenstates is orthonormal
\begin{align} \label{eq:k_orthogonality}
		&\langle u_{n q } | u_{m q} \rangle =  \int_0^L \diff x\, \langle u_{nq} |x\rangle \langle x |u_{mq} \rangle  = \delta_{nm} ,
\end{align}
and complete
\begin{align} \label{eq:completness_u}
	\sum_{n} 	| u_{n q} \rangle \langle u_{n q } | = \mathbbm{1},
\end{align}
for a given wave vector  $q$.
From the completeness relation in real space 
\begin{align} \label{eq:delta1}
	\delta (x-x_0) =  \sum_{n}  \sum_{q \in \text{BZ}} \langle x |u_{nq} \rangle \langle u_{nq} | x_0 \rangle ,
\end{align} 
we infer $\langle x |x_0  \rangle =  \delta (x_0 - x) $. 
From \cref{eq:k_orthogonality} the (over-) completeness relation for the basis states $|x \rangle$ follows  
\begin{align}
	 \int_0^{L} \diff x\,  |x  \rangle  \langle x |= \mathbbm{1}.
\end{align}
For the actual computation of the eigenfunctions we use the Fourier modes as  orthonormal basis  $\{|\nu\rangle: \nu \in \mathbb{Z}\}$ in $\mathcal{H}$ with  real-space representation 
$\langle x| \nu\rangle = \exp{(i Q_\nu x )} /\sqrt{L} $. 
It is favorable to express the Bloch functions $u_{nq}(x)$ in terms of their Fourier decomposition, and we express our eigenmodes in Dirac notation 
\begin{align} \label{eq:Fourierexpansion_p} 
	u_{nq} (x) &=   \langle x |u_{nq} \rangle   = \sum_{\nu \in \mathbb{Z}}  \langle x |\nu \rangle \langle \nu | u_{nq}  \rangle , 
\end{align} 
where the corresponding Fourier coefficients are provided by the integral 
\begin{align}
   \langle \nu | u_{nq} \rangle  =\int_0^L\!   \frac{\diff x}{\sqrt{L}}  e^{-i Q_\nu x } \, u_{nq}(x)  .
\end{align}
This is possible because all Bloch functions are lattice periodic.

 \subsection{Intermediate scattering function: definition and properties}
  We aim to analyze the characteristic function of the random displacements $\Delta x(t)\coloneqq x(t)- x(0)$, which corresponds to the self-ISF and provides full spatio-temporal resolution of the particle dynamics. Making use of the periodicity of the system, we introduce the Bravais lattice, defined as $\Lambda \coloneq \{nL :\, n \in \mathbb{Z} \}$. The reciprocal lattice is similarly defined as $\Lambda^* \coloneq \left\{Q_\mu=(2 \pi \mu / L) :\, \mu \in \mathbb{Z}\right\}$.
 Any wave vector $k$ can be uniquely decomposed as $k=q+Q_\mu$, where  $q \in \mathrm{BZ}$ lies within the first Brillouin zone (BZ) and $\mathrm{Q}_\mu \in \Lambda^*$ is a reciprocal lattice vector. 
We define the generalized ISF
\begin{align}\label{eq:ISF_def}
F_{\mu\nu}(q,t)\coloneq \langle e^{ -i(q+Q_\mu)x(t) } e^{i(q+Q_\nu)x(0)} \rangle,
\end{align}
with the \emph{mode indices} $\mu,\nu\in \mathbb{Z}$. In the following we refer to this quantity as the ISF. The brackets $\langle \cdots \rangle$ indicate an ensemble average. The conventional ISF to wave number $q+Q_\mu$ then corresponds to $F(q+Q_\mu,t) \coloneq
F_{\mu mu}(q,t)$ and probes diagonal correlations in reciprocal space. The off-diagonal elements $\mu \neq \nu$ encode \emph{Umklapp}-processes where wave vectors differ by a reciprocal lattice vector $Q_{\nu}-Q_{\mu}$. These Umklapp-processes enter by scattering from the periodic modulation, while they are absent in the homogeneous case. Thus, $F_{\mu\nu}(q,t)$ corresponds to a matrix-valued correlation function:
For a set of complex numbers $b_\mu(q) \in \mathbb{C}$ the weighted sum 
\begin{align} \label{eq:weightedoffdiagonal}
\langle b(q,t)^* b(q,0)\rangle=\sum_{\mu,\nu \in \mathbb{Z}} b_\mu(q)^* F_{\mu\nu}(q,t) b_\nu(q) ,
\end{align}
forms an autocorrelation function, with $b(q,t) = \sum_\mu b_\mu(q) e^{i(q-Q_\mu)x(t)}$. In particular, autocorreltaion functions display non-negative spectra\cite{hansen2013theory}. 
Notably, the diagonal elements, $F_{\mu\mu}(q,t)$, are autocorrelation functions.  For the case of purely relaxational dynamics, for example, Smoluchowski dynamics, autocorrelation functions are \textit{completely monotone}, i.e., all time derivatives exist and satisfy the inequality $(-1)^m \partial_t^m \langle b(q,t)^* b(q,0)\rangle \geq 0$ for $m \in \mathbb{N}_0$ and $t > 0$ \cite{lang2013mode}, ensuring monotonically decaying, non-oscillatory behavior.
In contrast, the individual off-diagonal elements, $\mu \neq \nu$, can exhibit non-monotonic behavior, including local minima or maxima. 

Further, we note that the wave vector $q$ is identical in both exponentials and that the diagonal elements, $F_{\mu\mu}(q,t)$, correspond to the conventional ISF evaluated at wave vector $q+ Q_\mu$. In translationally invariant systems only the diagonal elements are non-vanishing, since shifting the trajectory of a particle by an arbitrary displacement leads to an equally likely trajectory. 
For our case, the discrete symmetry is reflected in the sense that a common shift $x(t) \mapsto x(t) +R$  for all times by a lattice vector  $R \in \Lambda$ leaves the ISF invariant. 

Using the conditional probability density, the ISF of  \cref{eq:ISF_def} can be expressed as
\begin{align} \label{eq:ISF_integral}
		F_{\mu\nu}(q,t) =& \int_0^{NL} \! \diff x \int_0^{L} \! \diff x_0 \, e^{-i(q+Q_\mu)x } e^{i(q+Q_\nu)x_0} \nonumber \\
		&\times \mathbb{P}(x, t | x_0 )  p^{\text{eq}}(x_0) .
\end{align}
Here, we used that without restriction, the initial position of the particle can be chosen to be in a definite cell and is sampled from the equilibrium distribution for this single cell.

Reversely, we can compute the probability density by the backward Fourier transform
\begin{align}\label{eq:prob_by_ISF}
	&\mathbb{P}(x, t | x_0 ) p^\text{eq}(x_0) \nonumber \\ 
	& = \frac{1 }{L^2N} \sum_{\mu, \nu \in \mathbb{Z}} \sum_{q \in \text{BZ}}   F_{\mu\nu}(q,t)e^{ i(q+Q_\mu)x } e^{-i(q+Q_\nu)x_0} .
\end{align}
see  \cref{app:ISF_simplification} for the derivation. The previous relation also reveals that the conventional ISF is not sufficient to reconstruct the full probability density unless the system is fully translationally invariant. 

The explicit form of \cref{eq:ISF_integral} in terms of the Bloch functions is obtained by inserting \cref{eq:probability_density_S} and simplifies to express the stationary solution in terms of the eigenfunctions of the Schr\"odinger operator, using \cref{eq:eigenfunctions_0_S}. After some algebra and rearranging the terms, we find the final expression
\begin{align} \label{eq:ISF_integral_short}
	F_{\mu \nu}(q,t) =& \sum_{n }  e^{-\lambda_{nq}t } \left[  \int_0^{L} \! \diff x \, e^{ -iQ_\mu x }  u_{nq}(x) u_{00}(x)^{*} \right]  \nonumber \\
&\times \left[  \int_0^{L} \! \diff x_0\, e^{ -iQ_\nu x_0 }  u_{nq}(x_0) u_{00}(x_0)^{*} \right]^*,
\end{align}
see \cref{app:ISF_simplification} for more details.

To determine the functions $u_{nq}(x) $ we introduce the operator $\mathcal{L}_q $ for which the eigenvalue equation 
\begin{align} 
	\mathcal{L}_q   u_{nq}(x) = - \lambda_{nq}   u_{nq}(x),
\end{align}
holds. Straightforward substitution leads to  
\begin{align} \label{eq:operatorLk}
\mathcal{L}_{q}\coloneq & \frac{D \pi^2}{L^2} [ 2 u \cos( Q_1 x) - u^2 \sin^2(Q_1 x) ]  +D \partial_x^2 \nonumber \\
&+2 iq D  \partial_x  -D  q^2,
\end{align}
which can be abbreviated as $\mathcal{L}_{q} = \mathcal{L}_0 + \delta \mathcal{L}_{q} $. Here, the operator $\mathcal{L}_0$ (first three terms) encodes the interaction of the diffusive particle with the potential and the $q$-dependent operator  $\delta \mathcal{L}_{q}$ (last two terms) contains a drift or advection-like term linear, in the wave vector, and a diffusion-like term proportional to $q^2$. 

The matrix representation of the operator, is given by
 \begin{align}\label{eq:matrix_elements} 
\langle \mu | \mathcal{L}_0 | \nu \rangle  =&  \int_0^{L} \! \frac{\diff x}{L} e^{-i Q_\mu x} \mathcal{L}_0 e^{i Q_\nu x} =\frac{D\pi^2  }{L^2} \Big[   u (\delta_{\mu,\nu+1}+\delta_{\mu,\nu-1}) \nonumber \\
		   &+   \frac{ u^2}{4}  (\delta_{\mu,\nu+2}-2\delta_{\mu,\nu}+\delta_{\mu,\nu-2}) - 4  \mu^2 \delta_{\mu,\nu} \big] ,
\end{align}
and  
\begin{align}
	\begin{aligned} \label{eq:deltaL_k_matrix}
		\langle \mu |\delta \mathcal{L}_q |\nu \rangle  &=   ( - 4\pi q D  \nu/L   -D q^2)\delta_{\mu,\nu} .
	\end{aligned}
\end{align}
 The matrix $\mathcal{L}_0$  is a Hermitian matrix and pentadiagonal in the Fourier basis, i.e., it has non-zero elements only on the main diagonal and the two diagonals above and below it. The $q$-dependent matrix is diagonal in the Fourier basis. 
The eigenvalue problem $\mathcal{L}_q |u_{nq} \rangle = -\lambda_{nq} |u_{nq} \rangle$ is  computed numerically by diagonalizing the (truncated) Hermitian matrix 
\begin{align} \label{eq:operator_deltaLk}
	\sum_{\nu \in \mathbb{Z}} \langle \mu | \mathcal{L}_q | \nu \rangle \langle \nu | u_{nq}  \rangle &= -\lambda_{nq} \langle \mu | u_{nq}  \rangle. \end{align}

\begin{figure}[!ht]
    \centering
    \includegraphics[width=\linewidth]{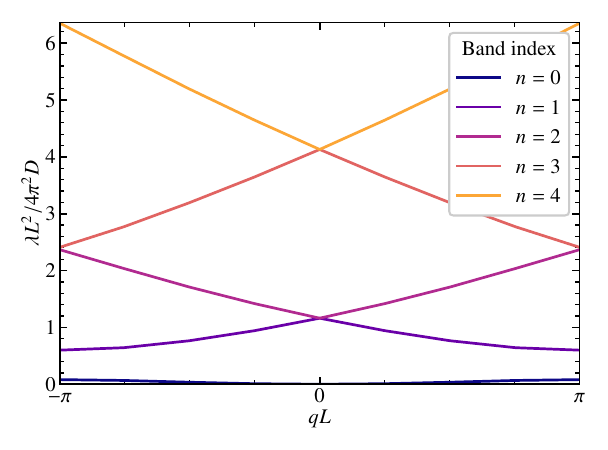}
    \caption{Eigenvalues $\lambda_{nq}$ of the operator $\mathcal{L}_q$ for $U_1=1.0\,k_BT$  are shown, for the five lowest bands in the first Brillouin zone. }
    \label{fig:Fig2}
\end{figure}

The time evolution of the ISF is encoded in the eigenvalues and eigenfunctions of the operator $\mathcal{L}_q$. The eigenvalues $\lambda_{nq}$ form continuous bands as the number of cells goes to infinity, $N\to\infty$, see   \cref{fig:Fig2}. All eigenvalues are non-negative, and the only zero eigenvalue is in the center of the Brillouin zone at the lowest band. Only the lowest bands are significantly affected by the potential $U(x)$. For $\lambda \gtrsim\pi^2  D u/L^2$ the bands correspond to a particle freely diffusing without underlying potential modulation.  
Albeit the bands come very close at the edges of the Brillouin zone, we checked numerically that they do not touch. For symmetric potentials the avoided crossing theorem does not apply and in principle bands can cross. For the simple cosine potential, one can actually show that \emph{all} eigenvalues for $q=0$ except for $\lambda_{00}=0$ are twofold degenerate, see \cref{app:orthonormality_simplification}.

Finally, the ISF in \cref{eq:ISF_integral_short} can be conveniently expressed in a spectral representation using the Fourier basis
\begin{align} \label{eq:ISF_braket}
	F_{\mu \nu}(q,t) 		= &\sum_{n}  e^{-\lambda_{nq}t}  \\
	&\times \sum_{\sigma, \tau \in \mathbb{Z}} \langle u_{00}|\sigma \rangle  \langle \sigma+\mu | u_{nq}  \rangle \langle u_{nq}|\tau+\nu  \rangle    \langle \tau | u_{00}  \rangle  ,  \nonumber
\end{align}
see \cref{app:ISF_simplification} for the derivation. The previous relation reveals that $\sum_{\mu,\nu} b_\mu(q) ^*F_{\mu\nu}(q,t)b_\nu(q) $ is completely monotone.

 Of particular interest is the conventional ISF with wave vector in the first Brillouin zone, $F(q,t)\coloneqq~F_{00}(q,t)$,
 which simplifies upon exploiting the completeness of the Fourier basis
\begin{align} \label{eq:ISF_braket_0}
	F(q,t) 
		&=  \sum_{n}  e^{-\lambda_{nq}t} \langle u_{00}| u_{nq}  \rangle \langle u_{nq}| u_{00}  \rangle =  \langle u_{00}| e^{\mathcal{L}_q t} | u_{00}  \rangle .
\end{align}

Since all eigenvalues are strictly larger than zero, except for $\lambda_{00}=0$ in the lowest band and in the center of the Brillouin zone, $F(q,t)$ decays to zero for large time, $t\to 0$ for $q\neq 0$.

\subsubsection{Symmetry relations}
The symmetries of the generalized ISF are derived following the same arguments as in Ref.~\cite{Lang2012}. 
For equilibrium dynamics, time inversion symmetry holds and the ISF is even in time. Time-translation invariance  then reveals the relations
\begin{align}\label{eq:time_inversion}
F_{\mu\nu}(q,t) = F_{\mu\nu}(q,-t) = F_{\nu\mu}(q,t)^{*} .
\end{align}
For symmetric potentials $U(-x) = U(x)$, space-inversion symmetry implies 
\begin{align}\label{eq:space_inversion}
F_{\mu\nu}(q,t) = F_{\mu\nu}(q,t)^{*} = F_{-\mu, -\nu}(-q,t). 
\end{align}
In particular, the ISF is a real quantity and symmetric upon interchanging its mode indices. 
At the edge of the Brillouin zone we have the additional relation
\begin{align}\label{eq:BZ_edge}
F_{\mu\nu}(\pi/L,t) = F_{\mu+1, \nu+1}(-\pi/L, t) = F_{-(\mu+1),-(\nu+1)}(\pi/L,t ) .
\end{align}
\subsubsection{Short- and long- time limits}
For the simple cosine potential, \cref{eq:cosine} , the short-time limit of the ISF can be calculated explicitly 
\begin{align}\label{eq:init_ISF}
F_{\mu\nu}(q,t=0) &= \langle \exp( -i Q_{\mu-\nu} x(0) ) \rangle = \int_0^L e^{- i Q_{\mu-\nu} x} p^{\text{eq}}(x) \diff x \nonumber \\
 &= (-1)^{\mu-\nu} \frac{I_{\mu-\nu}(u)}{I_0(u)}.   
\end{align}
From \cref{eq:ISF_integral_short} and  $u_{00}(x) = \sqrt{p^\text{eq}(x)}$, one infers that  for $q=0$ the ISF displays a non-vanishing   long-time limit 
   \begin{align}\label{eq:long_time_limit}
	   F_{\mu\nu}(0,t \to \infty) = \langle e^{-i Q_\mu x(t)} \rangle \langle e^{i Q_\nu x(0)}\rangle. 
\end{align} 
The factorization of the limit can be interpreted as the system being ergodic. 
For the simple cosine potential, \cref{eq:cosine}, the limit can be calculated explicitly 
\begin{align} \label{eq:ISF_braket_infity}
	F_{\mu\nu}(0,t \to \infty) 	&= \frac{(-1)^{\nu+\mu}\mathrm{I}_\mu(u)\mathrm{I}_\nu(u)}{\mathrm{I}_0^2(u)}
\end{align}
An equivalent formal expression for the long-time limit follows from \cref{eq:ISF_braket}
\begin{align}\label{eq:long_time_limit_explicit}
F_{\mu\nu}(0,t\to \infty) =& 
\left[ \sum_{\sigma \in \mathbb{Z} } \langle u_{00} | \sigma\rangle \langle \sigma + \mu | u_{00} \rangle \right] \nonumber \\
&\times \left[ \sum_{\tau\in \mathbb{Z}} 
 \langle u_{00} | \tau + \nu \rangle \langle \tau | u_{00} \rangle \right].
\end{align} 
Since $u_{00}(x) = \sqrt{p^\text{eq}(x)}$ and employing the Jacobi-Anger expansion~\cite{NIST:DLMF}, the Fourier coefficients $\langle \mu | u_{00} \rangle$ can be calculated explicitly
\begin{align} \label{eq:eigenvectors_q_0}
	\langle  \mu | u_{00} \rangle =
		\frac{ (-1)^{\mu} \mathrm{I}_\mu(u/2)}{\sqrt{ \mathrm{I}_0(u)}}  .
\end{align} 
With Neumann's addition theorem for modified Bessel functions~\cite{NIST:DLMF}  the sums in \cref{eq:long_time_limit_explicit} 
can be performed and we recover  \cref{eq:ISF_braket_infity}. Note that \cref{eq:init_ISF,eq:ISF_braket_infity} represent static quantities that only depend on the potential amplitude.

\subsection{Mean-squared displacement and non-Gaussian parameter}
The goal of this subsection is to elaborate the low-order moments of the fluctuating variable $\Delta x(t)$. They can be derived from the conventional ISF,  $F(q,t)=F_{00}(q,t)$, which is the characteristic function of the random displacements. As moments of odd order vanish by symmetry, we compute only the even moments. The lowest nontrivial is the time-dependent mean-squared displacement $\langle [ \Delta x(t) ]^2\rangle$. We also define  the time-dependent diffusion coefficient via the derivative of the mean-squared displacement (MSD)
\begin{align} \label{eq:diffusivity}
	D(t) \coloneqq  \frac{1}{2} \frac{\diff \langle [ \Delta x(t) ]^2\rangle }{\diff t}. 
\end{align}
 The long-time dynamics is diffusive, in particular, the long-time limit $D_\infty \coloneqq  D(t\to \infty)>0$ is finite and defines the long-time diffusion constant.  An analytic expression for arbitrary periodic potentials is known~\cite{Lifson_1962,Festa_1978,Reimann_2002,Straube_2024,Spiechowicz_2023}. In particular, for the simple cosine potential it evaluates to 
\begin{align} \label{eq:diffusivity_long_time}
	D_\infty= \frac{D}{I_{0}^{2}(u)}= 2\pi D e^{-2u}[u+O(1)] \qquad \text{for } u \to\infty . 
\end{align}
The exponential suppression of the diffusion constant reflects Kramers's rule for hopping over a potential barrier\cite{risken1996fokker}.
A convenient measure to discuss deviations from simple diffusion is the non-Gaussian parameter \cite{Rahman,hansen2013theory}
\begin{align} \label{eq:non_Gaussian}
	\alpha_2[\Delta x(t)  ] \coloneqq \frac{\langle [ \Delta x(t) ]^4\rangle }{3 \langle [ \Delta x(t) ]^2\rangle^2} -1 .
\end{align}
The derivation for extracting the moments from the ISF is similar to Refs. \citenum{Chepizhko_2022,Rusch_2024}, with only the essential steps presented here. The key idea is to apply perturbation theory for small wave vectors $q$ and compare it to the Taylor series of the ISF, which yields the moments
\begin{align} \label{eq:moment_expansion}
	F(q,t) &= 1 - \frac{q^2}{2} \langle [\Delta x(t)]^2 \rangle   + \frac{q^4}{4!} \langle [\Delta x(t)]^4 \rangle  + \cdots,
\end{align}
where we already exploited that all odd-order moments vanish. 
The ISF, \cref{eq:ISF_braket_0}, will be expanded in powers of $q$ by  $\mathcal{L}_q = \mathcal{L}_0 - D q^2 + \delta \mathcal{L}_q^1$ with 
$\delta \mathcal{L}_q^1 = 2 i q D \partial_x $. We rely on the Dyson representation
\begin{align} \label{eq:dyson_representation}
	&e^{(\mathcal{L}_0+\delta \mathcal{L}_{q}^1 ) t } = e^{\mathcal{L}_0 t} + \int_0^t \!\diff s \, e^{\mathcal{L}_0 (t-s)} \delta \mathcal{L}_{q}^1 e^{(\mathcal{L}_0 +\delta \mathcal{L}_q^1 ) s} .
\end{align} 
Replacing the time-evolution operator in the integral on the right-hand side iteratively generates the Born series, see Ref.~\citenum{Rusch_2024}. The main simplification steps are to make use of the fact that $e^{\mathcal{L}_0 t}  | u_{n0} \rangle = | u_{n0} \rangle $ and also   $  \langle u_{n0} | e^{\mathcal{L}_0 t} = \langle  u_{n0} | $ and to insert complete basis sets, \cref{eq:completness_u}, for $q=0$. Occurring integrals can be formally evaluated and the terms are simplified to obtain the final result, which is similar to the result in Ref.~\citenum{Rusch_2024}, but slightly changed for our operator and eigenvectors. We find the formal expression
\begin{align}\label{ISF_low_moments}
F(q,t) 	=& e^{-D q^2 t} \langle u_{00}|e^{(\mathcal{L}_0+\delta \mathcal{L}_{q}^1 ) t }  | u_{00}  \rangle \nonumber \\
  =& e^{-Dq^2 t} [ 1+ \frac{q^2}{2} \tilde{F}_2(t) + \frac{q^4}{4!} \tilde{F}_4(t) + O(q^6) ] ,
\end{align}
with 

\begin{align}
\tilde{F}_2(t)  =& \frac{2}{q^2} \sum_{n } \frac{e^{-\lambda_{n0} t} + \lambda_{n0} t -1}{\lambda_{n0}^2} 
	\langle u_{00} |\delta \mathcal{L}_{q}^1 | u_{n0} \rangle \langle u_{n0} |  \delta\mathcal{L}_{q}^1 |  u_{00} \rangle \nonumber
\end{align}
and,

\begin{widetext}
\begin{align}
    \tilde{F}_4(t) =& \frac{24}{q^4} \sum_{n} \sum_{m} \sum_{p} 
	\left[\frac{e^{-\lambda_{n0} t}+\lambda_{n0}  t-1}{\lambda_{n0}^2 (\lambda_{n0} -\lambda_{m0} ) (\lambda_{n0} -\lambda_{p0} )}\right. + \left. \frac{e^{-\lambda_{m0} t}+\lambda_{m0}  t-1}{\lambda_{m0}^2 (\lambda_{m0} -\lambda_{n0} ) (\lambda_{m0} -\lambda_{p0} )}+\frac{e^{-\lambda_{p0} t}+\lambda_{p0}  t-1}{\lambda_{p0} ^2 (\lambda_{p0} -\lambda_{m0} ) (\lambda_{p0} -\lambda_{n0} ) } \right] \nonumber \\
	&\times \langle u_{00} | \delta \mathcal{L}_{q}^1  |     u_{n0} \rangle \langle u_{n0} | \delta \mathcal{L}_{q}^1 | u_{m0} \rangle \langle u_{m0} | \delta \mathcal{L}_{q}^1 |     u_{p0} \rangle \langle u_{p0} | \delta \mathcal{L}_{q}^1 | u_{00} \rangle.
\end{align}
\end{widetext}
Here, all sums over $n, m, p$ include all bands and therefore formally the expression causes divisions by zero if a band index corresponds to the lowest band or two band indices correspond to the same band. In both cases the corresponding numerators also vanish. The appearance of the zero divisors can be avoided in the first place  
by handling these case separately before performing the integrals in the simplification steps. Here we follow a different route to keep the expressions simple by analytically continuing the expression for the case of zero numerators/denominators. 

 A further complication arises in the case of a simple cosine potential, since all eigenvalues  $\lambda_{n0}$, additionally,  are twofold degenerate, except for the ground state causing additional zero divisors. However, as in degenerate perturbation theory, one can choose basis states such that the matrix elements of $\delta \mathcal{L}_q^1$ coupling different states to the same eigenvalues vanish. Since $\mathcal{L}_q^1$ anticommutes with space inversion, only states of different parity couple, however, because $u_{n0}$ are either even or odd, no zero divisors occur. 
 
The low-order cumulants of the random variable $\Delta x(t)$ are generated  upon expanding the logarithm of the ISF in powers of the wave vector $q$
\begin{align}
\ln F(q,t) =& - \frac{q^2}{2} \langle [\Delta x(t)]^2 + \frac{q^4}{4!} \{ \langle [ \Delta x(t)]^4 \rangle \nonumber \\
& - 3\langle [\Delta x(t) ]^2 \rangle^2 \} + O(q^6)  .
\end{align} 
To order $O(q^2)$ we find the mean-square displacement as first nonvanishing cumulant
\begin{align}\label{eq:msd}
\langle [ \Delta x(t) ]^2 \rangle =& 2Dt - \frac{2}{q^2} \sum_{n \neq 0} \frac{e^{-\lambda_{n0} t} + \lambda_{n0} t -1}{\lambda_{n0}^2} \nonumber \\
&\times \langle u_{00} | \delta \mathcal{L}_{q}^1 | u_{n0} \rangle  \langle u_{n0} | \delta \mathcal{L}_{q}^1 |  u_{00} \rangle  ,
\end{align}
where no contributions from $n=0$ as intermediate state arises since the transition matrix element vanishes
\begin{align}\label{eq:no_drift}
\langle u_{00} | \delta \mathcal{L}^1_q | u_{00} \rangle = 2i q D  \int_0^L u_{00}(x)^* \partial_x u_{00}(x)  = 0, 
\end{align}
where the last equality follows by integration by parts and observing that $u_{00}(x)$ is real.
The time-dependent diffusion coefficient, \cref{eq:diffusivity}, can be computed using the time-derivative of \cref {eq:msd}. 
\begin{align} \label{eq:diffusivity_explicitely}
	D(t) =& D+\frac{1}{ q^2 }    \sum_{n \neq 0} \frac{ e^{-\lambda_{n0} t} -1 }{\lambda_{n0}}  \langle u_{00} | \delta \mathcal{L}_{q}^1 | u_{n0} \rangle 
	 \langle u_{n0} | \delta \mathcal{L}_{q}^1 |  u_{00} \rangle .
\end{align}
For the fourth cumulant we need to collect terms of order $O(q^4)$ and we obtain 
\begin{align}
& \langle [ \Delta x(t)]^4 \rangle - 3\langle [\Delta x(t) ]^2 \rangle^2  =  \tilde{F}_4(t) - 3 [ \tilde{F}_2(t)]^2 .
\end{align}

For completeness, let us argue explicitly that all odd powers in $q$ in the expansion of $F(q,t)$, \cref{ISF_low_moments}, vanish. A term linear in $q$ corresponding to a mean drift would involve the matrix element $\langle u_{00} | \delta \mathcal{L}^1_q | u_{00} \rangle$ which is shown to vanish in \cref{eq:no_drift}.  
This vanishing of the mean drift is, of course, a general property in equilibrium. For any symmetric potential all odd moments vanish. The expansion of $F(q,t)$ in \cref{ISF_low_moments} generates a chain $\langle u_{00} | \delta \mathcal{L}_q^1 | u_{n0} \rangle \langle u_{n0} | .... | u_{p0} \rangle \langle u_{p0} | \delta \mathcal{L}_q^1 | u_{00} \rangle$ of products of matrix elements. By parity and the property of the operator $\delta \mathcal{L}_q^1$ a matrix element is non-vanishing only if the states are of different parity. Thus, for the chain to yield a non-vanishing contribution, the first intermediate state has to be odd, the second even, and so on. However, the last state is the ground state again which is even. Therefore, only even powers of $q$ are generated.

\section{Comparison of the theoretical framework with experimental and simulation results} \label{sec:Results}

In this section, we analyze the observables and compare the theoretical predictions with the experimental results. First, we examine the diagonal and off-diagonal elements of the ISF for various wave vectors and potential amplitudes. Next, we discuss the mean-square displacement (MSD), time-dependent diffusivity, and the non-Gaussian parameter for different potential amplitudes.  
To facilitate the comparison with the experimental results, we restore units; in particular, we use the amplitude of the modulation $U_1$, \cref{eq:cosine}, as a control parameter. The presented experimental results correspond to the mean of multiple measurements, with error bars indicating the standard deviation. Each measurement consisted of a sufficiently large number of recorded trajectories, from which the observable of interest was computed as an average. To enable comparison with theoretical predictions, we established the relationship between the potential amplitude $U_1$ and laser power (LP) (see \cref{app:Calibration} for details). Additionally, length and time scales were calibrated by determining the short-time diffusion coefficient $D$ and the characteristic period $L$ from the experimental data. We determined from the initial slope of the mean-square displacement an average short-time diffusion coefficient of $(0.050 \pm 0.002)$ \textmu m$^2$/s.

 \begin{figure}[ht!]
    \centering
    \includegraphics[width=\linewidth]{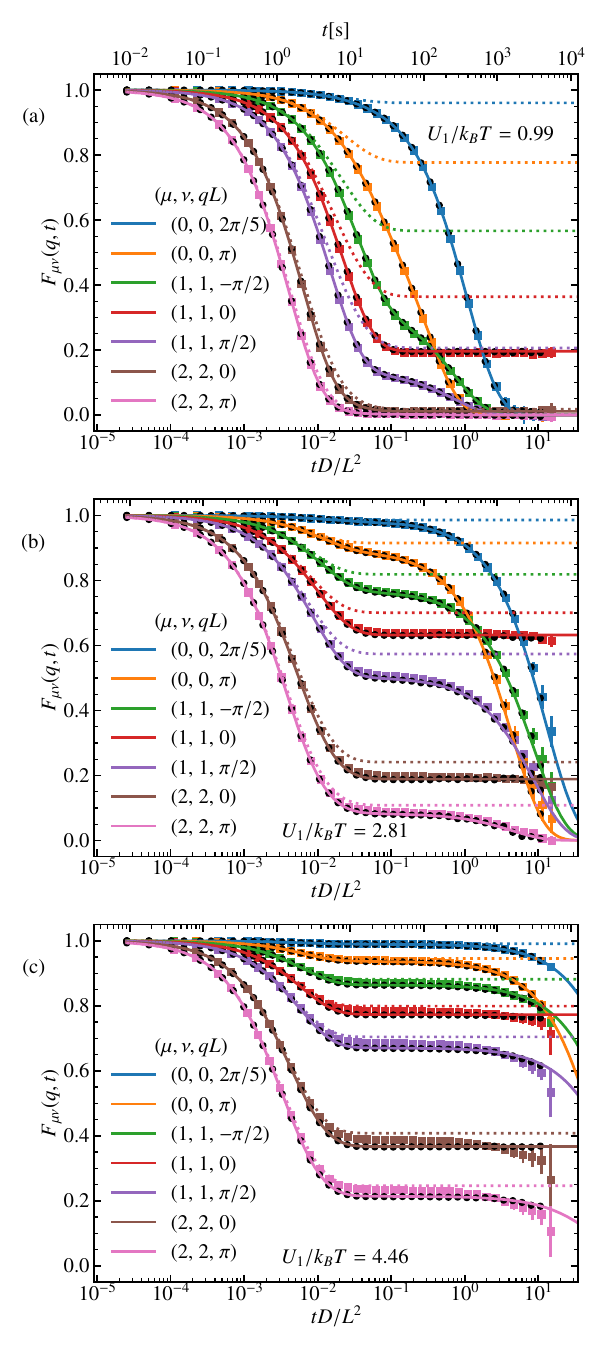}
    \caption{Diagonal ISF for various mode indices $\mu=\nu$ and  wave vectors $q$ and three different amplitudes $U_1$, (a-c). Full colored lines correspond to the theory and colored squares to the experimental results. Black circles represent the simulation results. The dotted lines represent the harmonic approximation. }
    \label{fig:Fig3}
\end{figure}

\subsection{Diagonal ISF, $\mu = \nu$}\label{sec:results_ISF}
 
The diagonal ISF is computed from our experimental data relying on the definition given in \cref{eq:ISF_def} for $\mu=\nu$. We have checked that its imaginary part is in deed negligible reflecting the mirror symmetry of the potential. We compare the experimental results to the numerical ones relying on the spectral representation in \cref{eq:ISF_braket}. Last, all results are corroborated by Brownian-dynamics simulations. 
The ISF  is displayed in \cref{fig:Fig3} for a range of different wave vectors $Q_\mu \in \Lambda^*$ and $q \in \text{BZ}$ for three distinct amplitudes $U_1$.

We first focus on the behavior for $q\neq 0 $, where all ISF eventually relax to zero. For moderate potential amplitudes $U_1 \approx k_BT$, the potential is not high enough to significantly inhibit hopping between different potential valleys, yielding a single-step relaxation. For larger amplitudes, a two-step process occurs.  The particle initially freely diffuses with short-time diffusion coefficient $D$ until the potential forces become dominant. For $U_1 \gtrsim k_BT$ the motion occurs essentially at the bottom of the potential, which can be approximated by a harmonic well 
\begin{align}
    U(x)\approx -U_1+ \frac{U_1 Q_1^2 (x-L/2)^2}{2}.
\end{align} 
The particle then locally equilibrates on the time scale of the harmonic relaxation time, $\tau=(L^2/ 4\pi^2D)(k_BT /U_1)$, and the ISF saturates at a plateau value, see \cref{fig:Fig3}. The ISF within the  harmonic approximation can be calculated explicitly, see \cref{app:HA}. 
For large barriers, \cref{fig:Fig3}\,(c) the harmonic approximation quantitatively describes the relaxation to the plateau value for wave vectors resolving smaller length scale than a period $L$.  
The relaxation from the plateau occurs on a much larger time scale provided by Kramers' theory  $\tau_\text{K}\propto \exp(2U_1/k_BT) $. Once the particle overcomes the barrier and reaches additional minima, the ISF eventually decays to zero.
For small wave vectors and long times, the hydrodynamic regime is reached $F_{00}(q,t)= \exp(-D_\infty q^2 t)$. In this regime, the wave vectors only resolve the motion over many periods at time scales much larger than Kramers' escape time.  
Our analytical predictions as well as the simulation results within the Smoluchowski picture of a simple cosine potential show excellent agreement with the experimental results. 

For the wave vector in the center of the BZ, $q=0$, the ISF does not decay to zero in the long-time limit  $t \to \infty$, rather approaches a finite value, as computed in \cref{eq:ISF_braket_infity}. In \cref{app:Calibration} we used this feature to calibrate the laser power of the experiment to the theoretical $k_B T$ value.

\subsection{Off-diagonal ISF, $\mu \neq \nu$}\label{sec:results_ISF_offdiag}
\begin{figure}[]
    \centering
    \includegraphics[width=\linewidth]{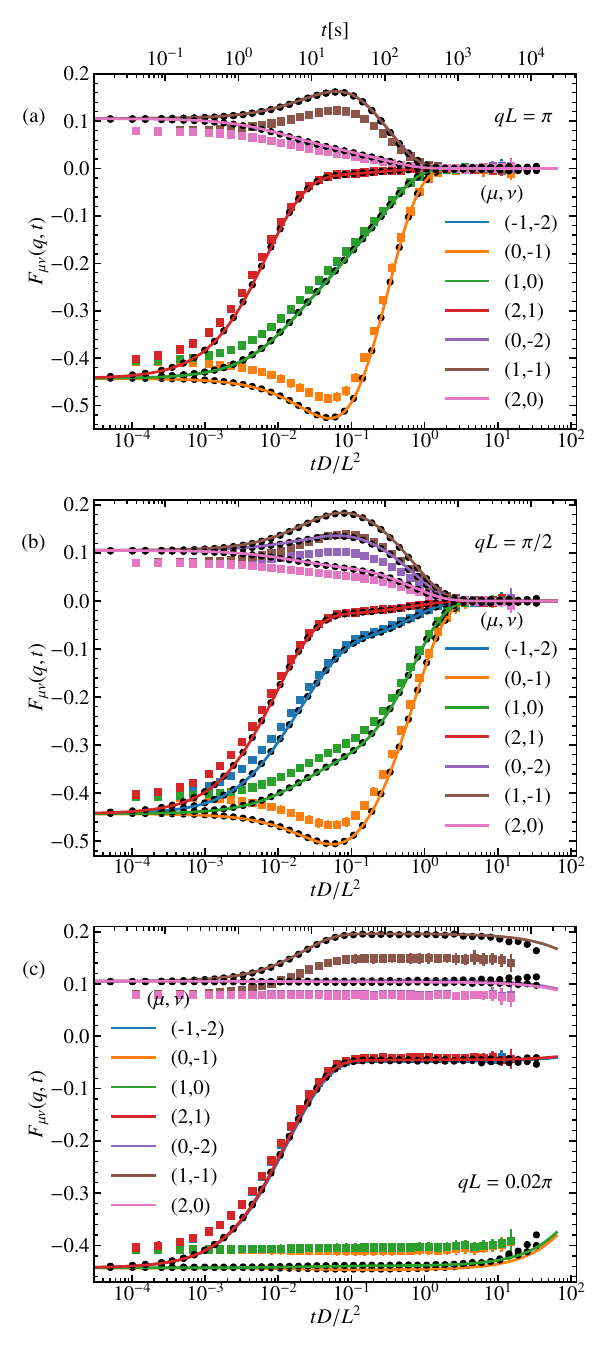}
    \caption{Off-diagonal ISF for $\nu\neq\mu$ and $U_1=0.99 k_BT$  for different mode indices $(\mu,\nu)$ and wave vectors $q$, (a-c). Full colored lines correspond to the theory and squares to the experimental results. Black circles represent the simulation results. In panel (a) at the edge of the Brillouin zone, $q=\pi/L$,  for $(\mu,\nu) = (1,-1)$ and $(0,-2)$ coincide, as well as the ones for $(1,0)$ and $(-1,-2)$.  In panel (c) for wave vectors $q \to 0$ the curves of $(\mu,\nu)$ and  $(-\nu,-\mu)$ approach each other, where $\mu,\nu \in \mathbb{Z}$.}\label{fig:Fig4}
\end{figure}

For the off-diagonal ISF, $\mu\neq \nu$, we present results for different modes corresponding to $|\mu-\nu|=1$ and $|\mu-\nu|=2$. The results are shown for three different wave vectors $q$ and the amplitude $U_1=0.99 k_BT$, see \cref{fig:Fig4}. The general symmetries of the ISF for symmetric potentials are summarized in \cref{eq:time_inversion,eq:space_inversion}.

For wave vectors at the edge of the BZ  $q = \pi/L$ additional symmetries of the ISF, \cref{eq:BZ_edge}, hold. For example, the curves for $(\mu, \nu)=(1,-1 )$ and $(0, -2)$, or $( 1, 0)$ and $(-1, -2)$ are identical. 
 
For $q=0$ both the initial value and the long-time limit are non-zero. As soon as $q\neq 0$ the curves decay to zero for long times. 
If $|\mu-\nu|$ is odd, the initial value is negative and for even differences the initial value of the ISF is positive, \cref{eq:init_ISF}. And also according to \cref{eq:ISF_braket_infity} for odd (or even) values, the long-time limit is negative (or positive, respectively). In contrast to the diagonal elements of the ISF, we find no longer strictly monotone behavior but minima and maxima, see \cref{eq:weightedoffdiagonal}.

Results from the experimental data of the off-diagonal ISF closely follow the shape of the analytical predictions and simulation results at long times. However, a clear deviation is observed at short and intermediate times, as shown in Fig. \ref{fig:Fig4}. We attribute these discrepancies to experimental factors not implemented in the theoretical model. We consider the primary factor to be the spatial inhomogeneities in the periodicity and amplitude of the light-induced potential across the field of view, which slightly deviates from the ideal cosine form assumed in the theoretical model. As a result, individual particles effectively experience slightly different periodicities and amplitudes across the field of view. A detailed characterization of the amplitude and periodicity in the field of view is shown in \cref{app:Characterization}. Additionally, confinement effects due to the  two-dimensional nature of the system and inertial effects not considered in the theoretical framework are factors that might also contribute to these discrepancies. Despite these differences, experimental results for the indices $\mu$ and $\nu$ with equal $|\mu-\nu|$ collapse into a common intercept (short-time limit), capturing the expected phenomenology from the theoretical predictions given by Eq. (\ref{eq:init_ISF}), though with a noticeable shift respect to theory. 

\begin{figure}[]
    \centering
    \includegraphics[width=\linewidth]{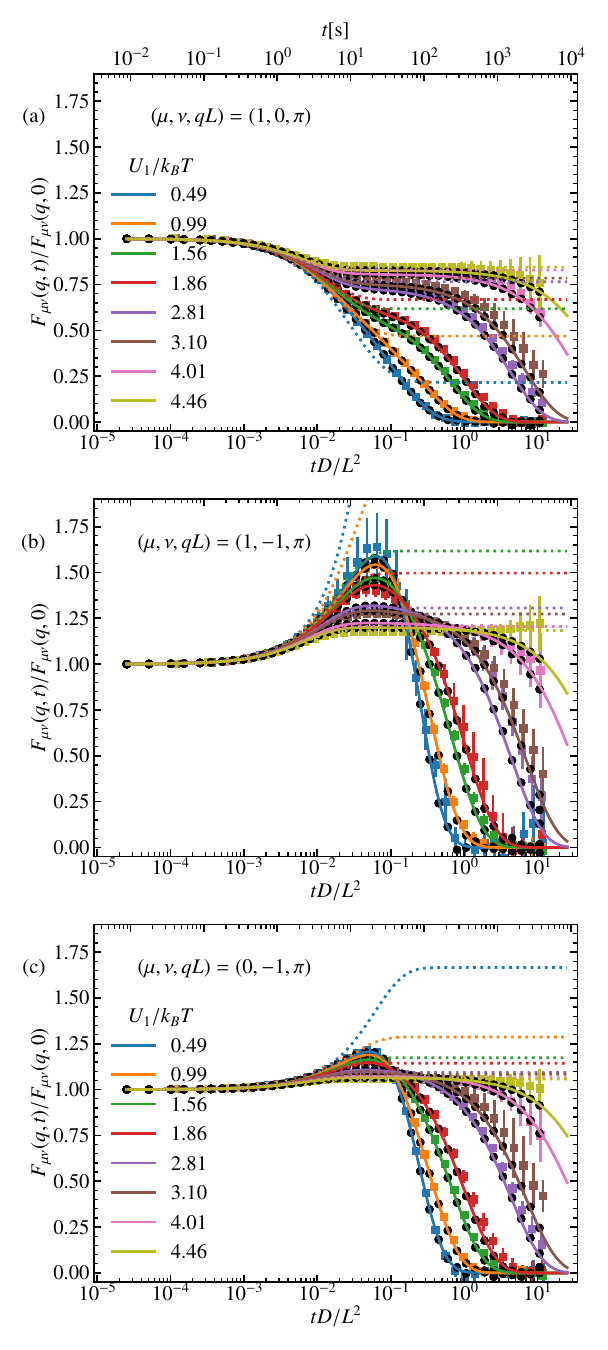}
    \caption{Off-diagonal ISF for different amplitudes $U_1$ for wave vector $q=\pi$ three different mode indices $(\mu,\nu)$ , (a-c). Full colored lines correspond to the theory and squares to the experimental results. Black circles represent the simulation results and dotted lines the harmonic approximation.} 
	\label{fig:Fig5}
\end{figure}

Theoretically, each Fourier component of the ISF is defined in terms of a well-defined lattice wave vector $2 \pi /L$. However, in experiments, achieving this level of precision is challenging because of variations in how individual particles interact with the optical field. To allow a more accurate comparison between theory and experiment, a normalization procedure is applied, an approach that is further examined in \cref{fig:Fig5}.
Figure \ref{fig:Fig5} illustrates the normalized ISF, $F_{\mu \nu}(q,t)/F_{\mu \nu}(q,0)$, for different potential amplitudes $U_1/k_BT$, three different $\mu,\nu$ combinations and, for all of them, $qL=\pi$. The choice of $qL=\pi$ is particularly insightful, as it balances, for the measurements time window, sensitivity to both free diffusion and potential-induced localization, providing a clear distinction between different transport regimes. As it can be seen in Fig.  \ref{fig:Fig5}, after normalization, the agreement between experimental results, the theoretical framework and computer simulations is remarkable, confirming that the model effectively captures full description of the system. 

The most striking feature in the off-diagonal elements is the emergence of a maximum at intermediate times, $\tau \ll t \ll \tau_\text{K}$ as seen in \cref{fig:Fig5} (b) and (c). For larger times, the curves decay to zero for $q\neq 0$. From the harmonic approximation we anticipate the development of a plateau, whose value can be determined from \cref{eq:ISF_HA_ratio}. If $(Q_\mu +q)(Q_\nu+q)<0$, the plateau corresponds to a maximum, which is nicely approached for large potential amplitudes, see \cref{fig:Fig5} (b) and (c). If $(Q_\mu +q)(Q_\nu+q)>0$, the curves look similar to the diagonal elements of the ISF, where a simple plateau emerges, see \cref{fig:Fig5} (a). 
The slowing down of the relaxation towards the plateau or maximum as the potential amplitude grows, is captured as well by $\tau \propto 1/U_1$. 

Finally, it is important to emphasize that, although experimental factors cause deviations at short and intermediate times, normalization effectively accounts for these variations, leading to excellent agreement between theory, simulations and experiments.

\subsection{Low-order moments of $\Delta x(t)$: MSD, diffusivity and non-Gaussian parameter} \label{sec:results_MSD}

\begin{figure}[htbp!]
   \centering
 \includegraphics[width=\linewidth]{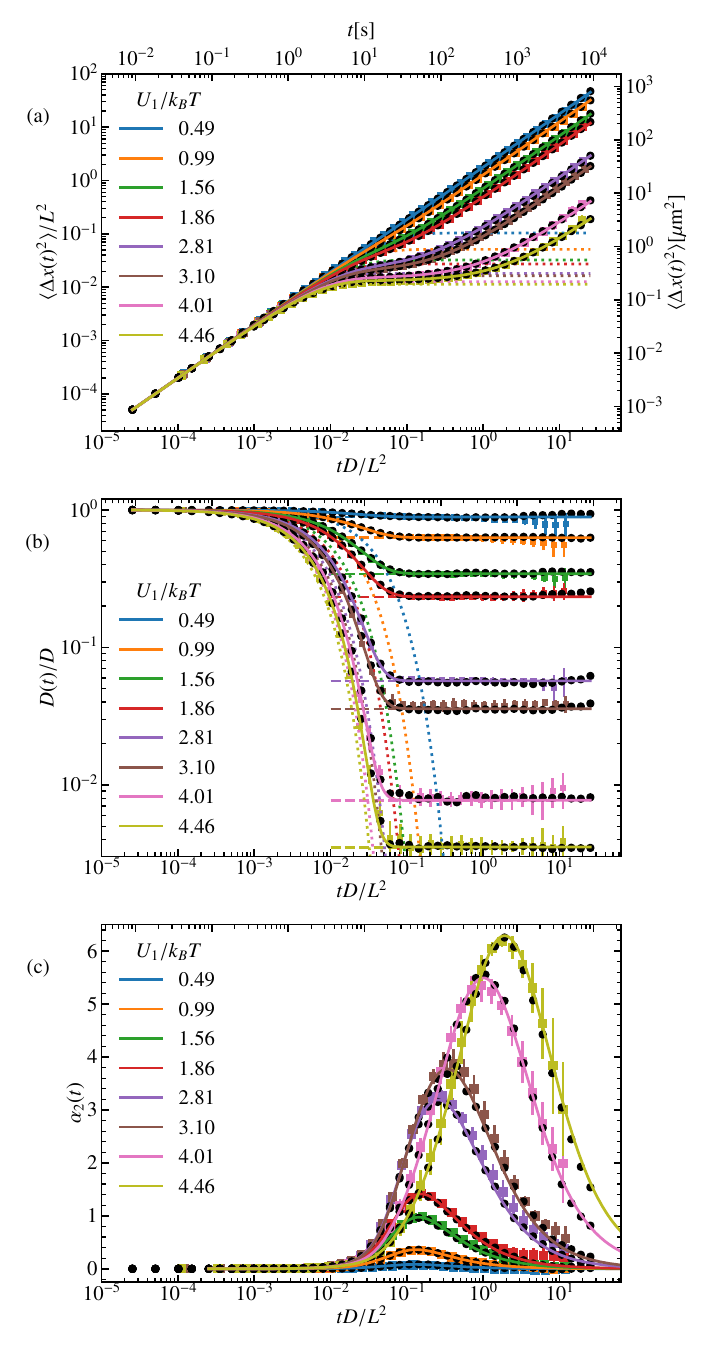}
   \caption{
   (a) MSD, (b) Time-dependent diffusion coefficient, and (c) non-Gaussian parameter for different potential amplitudes $U_1$. Full colored lines correspond to the theory and squares to the experimental results. Black circles represent the simulation results and dotted lines the harmonic approximation. The dashed lines in (b) display the theoretical long-term limit. }

	\label{fig:Fig6}
\end{figure}

The dynamics previously discussed in terms of the ISF, can also be found in the standard observables. Therefore, the discussion will be kept rather brief. 
The MSD, \cref{fig:Fig6}\,(a), is plotted for various amplitudes $U_1$ of the external potential. For short times, $t \ll (L^2/ 4\pi^2D)(k_BT /U_1)$,  with $U_1  \gtrsim k_BT$ or for times $t \ll L^2/ 4\pi^2D$ for very low amplitudes,  $U_1  \ll k_BT$, we observe the expected linear increase of the MSD, characteristic of free diffusion. For higher potential barriers, however, a plateau emerges around $t \approx (L^2/ 4\pi^2D)(k_BT /U_1)$, corresponding to the time scale at which the particle becomes temporarily trapped. At longer times, $\propto \exp(2 U_1/k_BT)$, the particle eventually overcomes the barrier and resumes diffusion, with  a reduced diffusion constant for higher amplitudes compared to lower ones. This behavior is further highlighted in the time-dependent diffusion coefficient, \cref{eq:diffusivity}. We find that the long-time diffusion coefficient decreases as $U_1$ increases, see \cref{fig:Fig6}\,(b). There we also report good agreement with the known values of the  long-time diffusion coefficient.
As expected, the harmonic approximation captures the behavior of the relaxation towards the plateau increasingly better for higher amplitudes.

Furthermore, we analyze the non-Gaussianity of the particle displacements using the parameter defined in \cref{eq:non_Gaussian}. As expected, for higher barriers, the particle dynamics become increasingly non-Gaussian, see \cref{fig:Fig6}\,(c). We observe that both very small and very large amplitudes pose challenges in experiments. For small amplitudes, it is difficult to distinguish the dynamics from those of a free particle, as the external potential has little effect. Conversely, for very large amplitudes, the low diffusivity makes it challenging to sample a sufficient number of particles that successfully hop over a barrier within the experimental observation time. 
However, the experimental results show excellent agreement with both simulations and analytical predictions. 

\section{Conclusion} \label{sec:conclusion}
In this work, we investigated the dynamics of dilute colloidal suspensions in the presence of a periodic potential. Our approach combined theoretical analysis, Brownian-dynamics simulations, and experimental measurements. 
By evaluating the spatio-temporal dynamics, we demonstrated that the behavior of a colloid in a periodic potential can be accurately described by analytical solutions of the Smoluchowski equation. This was achieved through the analysis of a \textit{generalized} ISF, which captures how particle positions are correlated in a periodic system. Low-order moments were derived, with a focus on the MSD, time-dependent diffusion coefficient, and non-Gaussian parameter. 

Based on the Smoluchowski equation reformulated in a Hermitian Schr\"odinger form, we found formal expressions in terms of a spectral-theoretical approach. The eigenfunctions were expressed in Bloch form, to make use of the periodic nature of the system. We found an analytic expression for the generalized ISF $F_{\mu \nu}(q,t)$. By using the time-dependent perturbation theory and Taylor expansion of the ISF we computed lower-order moments.
In our system, without memory effects, the ISF effectively captures the \textit{full} dynamics of colloidal particles in periodic potentials, revealing both short- and long-time diffusive behavior and trapping at intermediate times. 

We performed experiments on 2D dilute colloidal suspensions subjected to a periodic potential generated by two interfering laser beams. Using particle tracking, we obtained particle trajectories and averaged them to extract relevant observables. The laser power was calibrated to its corresponding amplitude value using two theoretical predictions (\cref{eq:diffusivity_long_time,eq:ISF_braket_infity}).
From our experimental data, we extracted the observables of interest and compared them to our analytical solutions and Brownian-dynamics simulations. We compared the results for various strengths of the amplitude of the potential and found excellent agreement between the theoretical description and experiments. 
The most sensitive observables were the off-diagonal elements of the generalized ISF, where slight differences in the experimental setup were amplified in the curves. It was crucial to ensure equivalent experimental conditions, minimizing variations in periodicity, potential amplitude, and confinement effects. To obtain good agreement with the analytical predictions, a normalization was necessary.

We have carried out a comprehensive test of the underlying fundamental dynamics of colloidal dynamics in a structured environment, combining theory, simulations, and experiments. By analyzing the generalized ISF, we identified new observables with distinct features, including a non-vanishing long-time limit. We provide a detailed theoretical framework and rationalize our findings through a harmonic approximation. Developing an explicit formula for the whole-time dependence of the ISF allowing for a comprehensive description of colloidal motion across all time scales. Furthermore, we introduced a new approach for calibrating the experiment using these observables, offering a reciprocal space alternative to conventional calibration methods. Comparing the results to a harmonic approximation, we confirm that the Brownian particle first diffuses freely, before it is temporarily trapped in the minima of the periodic cosine potential. For these times the dynamics are well approximated by a harmonic potential for large enough amplitudes, and only at longer times it hops over the potential barriers and once again exhibits diffusive behavior.

The analytical and experimental framework presented can be extended to more complex systems. Although this work focused on dilute suspensions, exploring more dense systems would allow us to study particle interactions and many-body effects. A possible other extension is the study of periodic lattices, where higher-dimensional effects and collective behavior become important. Investigating tilted washboard potentials could provide further insight into driven transport. Our framework is also applicable to a wide range of periodic systems beyond simple cosine potentials. Experimentally, the new observables could also be measured using differential dynamic microscopy.

\section*{Author Contributions}
MAES and TF conceived and administrated the project. YMS, ABZB, and MAES conducted the experimental work. 
RR and TF performed the analytic modeling. RR implemented the numerical analysis.  YMS, MAES, and RR performed computer simulations, analyzed the experimental data, and contributed to data visualization.
RR drafted the original manuscript, and all authors contributed to the manuscript.

\section*{Conflicts of interest}
There are no conflicts to declare.

\section*{Acknowledgments}
We dedicate this work to the memory of Stefan Egelhaaf, with sincere gratitude for his contributions and dedication to the field of soft matter.

We acknowledge the use of AI (ChatGPT, Grammarly, LanguageTool, Perplexity) for its assistance with grammar checking, translations, and text enhancement. 
This research was funded in part
by the Austrian Science Fund (FWF) 10.55776/I5257. For open access purposes, the author has applied a CC BY public copyright license to any author accepted manuscript version arising from this submission.
MAES acknowledges funding by Deutsche Forschungsgemeinschaft (DFG) - Project number 459399860 for financial support.

\begin{widetext}
	
	\appendix
\section{Appendix: Calibration} \label{app:Calibration}
\begin{figure}[htb]
    \centering
    \includegraphics[width=0.5\linewidth]{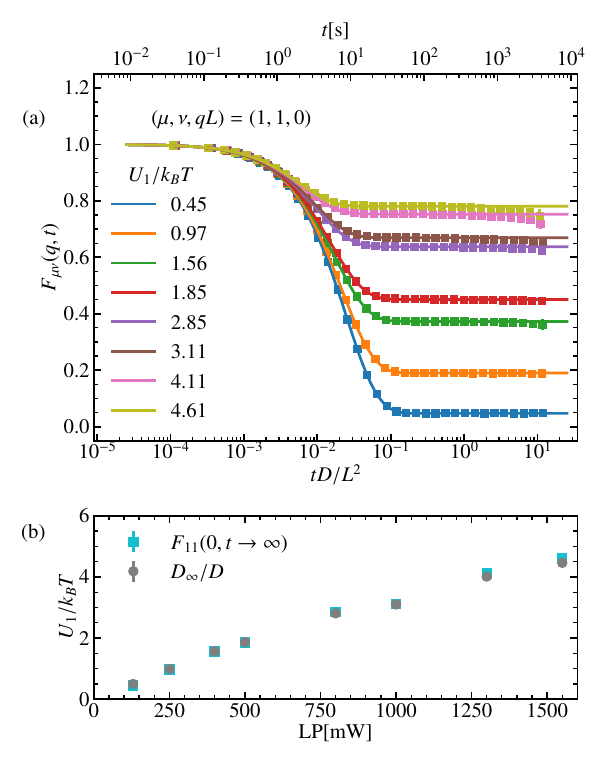}
    \caption{Panel (a) shows the ISF for the wave vectors $(\mu, \nu, qL) = (1,1,0)$ and for different amplitudes $U_1$. The markers represent experimental results, while the solid lines correspond to analytical predictions, with the amplitudes calibrated to match the experimental data. Panel (b) compares two different calibration methods. One based on the long-time diffusivity $D_\infty$ and the other on the long-time limit of the ISF, $F_{11}(0, t \to \infty)$. The laser powers (LP) used in the experiments, for which the calibration was performed, were 130mW, 250mW 400mW, 500mW, 800mW, 1000mW, 1300mW, and 1550 mW.}
	\label{fig:setupcal}
\end{figure}
The precise determination of the periodic potential amplitude ($U_1/k_BT$) as a function of laser power is a fundamental step in experiments involving colloidal particles confined in optical potentials. In this work, we perform both the traditional diffusion-based calibration and a novel approach utilizing the generalized intermediate scattering function (ISF).
The conventional approach to calibrating periodic potentials relies on measuring the normalized long-time diffusion coefficient ($D_{\infty}/D$) as a function of laser power. As shown in the central panel of \cref{fig:Fig6}, the diffusivity of particles decreases as the potential amplitude increases, since energy barriers progressively hinder diffusive motion. For a simple cosine potential, $U(x) = U_1 \cos(2 \pi x/L)$, the analytical relationship between the normalized diffusion coefficient and the potential amplitude is given by \cref{eq:diffusivity_long_time}.

Our alternative calibration method utilizes the generalized ISF, specifically its asymptotic behavior, which explicitly depends on the potential amplitude. For a simple cosine potential and $q=0$, the long-time limit of the generalized ISF is analytically described by \cref{eq:ISF_braket_infity}, where the dependence on the potential amplitude is evident. Specifically, for $\mu\!=\!\nu\!=\!1$, \cref{eq:ISF_braket_infity} ($F_{11}(0,t)$) evaluates the characteristic wave vector imposed on the system by the periodic potential, i.e., $Q_{\mu} \!=\! Q_{\nu}\! = \!2\pi/L$. In the top panel of \cref{fig:setupcal} we plot $F_{11}(0,t)$, which exhibits an opposite trend to diffusivity, with higher plateaus as laser power increases. It is important to note that this approach utilizes equilibrium correlation properties instead of transport characteristics, providing complementary insights into the system.

The calibration is performed for both methods by extracting plateau values as a function of laser power and solving \cref{eq:ISF_braket_infity,eq:diffusivity_long_time} for the diffusivity and generalized ISF methods, respectively. The calibration results from both methods are presented in \cref{fig:setupcal} (b), showing the relationship between laser power and dimensionless potential amplitude $U_1/k_BT$. The remarkable agreement between these independent approaches validates both the theoretical framework and the experimental implementation, as seen in the central panel of \cref{fig:Fig6} and \cref{fig:setupcal} (a). This strong consistency confirms the theoretical predictions and demonstrates the reliability of this method for calibration purposes.

\section{Appendix: Light-field spatial characterization} 
\label{app:Characterization}

This appendix provides a detailed look at the spatial characteristics of the laser-induced periodic potential, as observed through the motion of particles across the experimental field of view. 
To determine the spatial periodicity, we follow a similar approach as described in Ref.~\cite{Dieball2025}. However, for the estimation of the amplitude, we used a more direct estimation. Rather than fitting a sine wave to the derivative of the logarithmic particle density, we extracted the amplitude in each window by identifying its maximum and minimum values.
The extracted amplitude of the periodic potential as a function of the position along the field of view is shown in \cref{fig:Fig8}\,(a). The amplitude displays a clear spatial dependence, which arises from the Gaussian envelope of the beam~\cite{Jenkins_2008,Wei1998,Egelhaaf_2011}. The mean amplitude was measured to $U_1 = (1.03 \pm 0.08)\, k_BT$, with a coefficient of variation (CV) of 7.76\%. The inset histogram shows the distribution of the measured amplitude values.
\cref{fig:Fig8}\,(b) shows the spatial periodicity of the laser field in the same region. In contrast to the amplitude, the periodicity remains remarkably consistent throughout the field of view, as it is unaffected by the Gaussian beam envelope. It maintains an average value of $L = (4.15 \pm 0.07)\,\mu\mathrm{m}$, with a very low CV of 1.68\%. The inset histogram confirms this stability by showing a narrow, normally distributed spread of periodicities. The error bars in both figures represent variations between different experimental runs in the same spatial regions.

\begin{figure}[htb]
   \centering
 \includegraphics[width=0.5\linewidth]{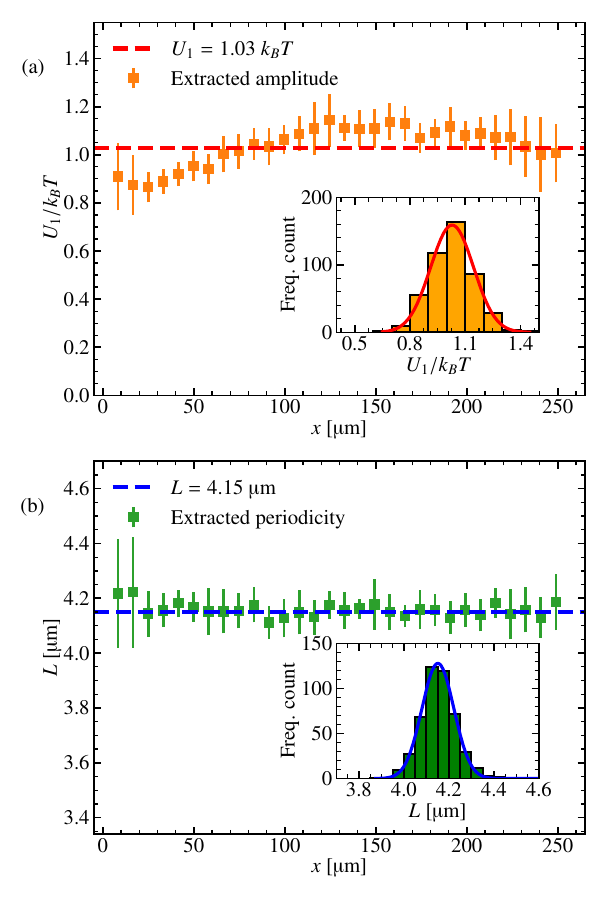}
   \caption{Panel (a): the spatial variation of the extracted potential amplitude $ U_1$ (orange squares) along the field of view, showing a mean value of  $U_1 = 1.03\,k_BT $(red dashed line) with a standard deviation of 0.08 (orange error bar). Inset: histogram of the extracted amplitude values fitted with a Gaussian distribution (red curve). 
   Panel (b): the spatial variation of the extracted periodicity $L$ (green squares), with a mean value of $L = 4.15$ \textmu m  (blue dashed line) with a standard deviation of 0.07 (green error bar). Inset: histogram of the extracted periodicities fitted with a Gaussian distribution (blue curve).
  }
	\label{fig:Fig8}
\end{figure}
\section{Appendix: Properties of the Bloch functions} \label{app:orthonormality_simplification}
In this appendix, we repeat the argument that the Bloch function $u_{nq}(x)$ to the same wave vector $q$ are orthonormal. Furthermore, we recall some properties of the Bloch functions, in particular, for the case of a symmetric potential.

\subsection*{Orthogonality of Bloch functions}
The overlap of two wave function can be simplified using the periodic Bloch functions and splitting the integral in cells
\begin{align} 
&	\langle \Psi_{nq^\prime}  | \Psi_{mq}  \rangle = \frac{1}{N} \int_0^{ N L }\! \diff x \,  e^{i (q-q^\prime) x} u_{nq^\prime} (x)^* u_{mq} (x)   =\frac{1}{N}  \sum_{j=0}^{N-1}\int_{ j L }^{(j+1) L }\! \diff x\,  e^{i (q-q^\prime) x}  u_{nq^\prime} (x)^* u_{mq} (x) \nonumber \\
		&= \frac{1}{N  }  \int_0^{L}\! \diff x\, e^{i (q-q^\prime) x} u_{nq^\prime} (x)^* u_{mq} (x) \sum_{j=0}^{N-1} e^{ i (q-q^\prime) j L }  .
		\end{align}
Since 
		\begin{align}
		\sum_{j=0}^{N-1} e^{ i (q-q^\prime) j L }  &= \begin{cases}
		\frac{\exp( i N( q-q^\prime) L  ) - 1 }{\exp( i (q-q^\prime) L  ) - 1} & \text{for } q\neq q^\prime \\
		N & \text{else}
		\end{cases} \nonumber \\
		& = N \delta_{q q^\prime}.
		\end{align}
for the discrete wave vectors in the Brillouin zone and because the wave functions are orthonormal $\langle \Psi_{nq^\prime}  | \Psi_{mq}  \rangle = \delta_{qq^\prime} \delta_{nm}$, Bloch functions to the same wave vector are orthogonal		
\begin{align*}
 \int_0^{L}\! \diff x\, u_{nq} (x)^* u_{mq} (x) = \delta_{nm} .
 \end{align*}

\subsection*{Symmetric band structure}
The Hermitian operator $\mathcal{L}_0$ is real. We introduce the eigenfunction $\Psi_\lambda(x)$ to  $\mathcal{L}_0 \Psi_\lambda(x) = -\lambda \Psi_\lambda(x)$. Taking the complex conjugate of both sides reveals that $\Psi_\lambda(x)^*$ is again eigenfunction to the same eigenvalue. Bloch theorem states that eigenfunctions can be expressed as $\Psi_{nq}(x) = \exp(i q x) u_{nq}(x)$ with $u_{nq}(x)$ periodic. Hence,  $u_{n, -q}(x)$ coincides with $ u_{nq}(x)^*$ up to  a phase factor. Without restriction, the phase factor can be chosen to be real. 
Therefore, the corresponding band structure is symmetric with respect to flipping the sign of the wave vector, $\lambda_{n, q} = \lambda_{n,-q}$.  Furthermore, the eigenfunctions in the center of the Brillouin zone are real $u_{n 0}(x) = u_{n 0}(x)^*$. 

\subsection*{Degeneracies for symmetric potentials}
We define the parity operator $\mathbf{P}$ acting on functions $\mathbf{P} \Psi(x) = \Psi(-x)$. For a symmetric potential $\mathbf{P} U(x) = U(-x) = U(x)$ and the parity operator commutes with $\mathcal{L}_0$. Then, with $\Psi_\lambda(x)$ eigenfunction, we find 
$\lambda \mathbf{P} \Psi(x)  = \mathbf{P} \mathcal{L}_0 \Psi_\lambda (x) = \mathcal{L}_0 \mathbf{P} \Psi_\lambda (x)$. 
Hence, $\mathbf{P} \Psi_\lambda(x) = \Psi_\lambda(-x)$ is again eigenfunction to the same eigenvalue. For the Bloch representation 
this implies $u_{nq}(-x) = \pm u_{n, -q}(x) $.  In particular for $q=0$ we find $u_{n0}(x)$ is either even or odd. 
As a consequence all matrix elements $\langle u_{m0} | \mathcal{L}_0 | u_{n0} \rangle$ vanish if the eigenfunctions $u_{m0}(x), u_{n0}(x)$ have different parity. In particular, the avoided crossing theorem does not apply, the eigenvalues at the center of the Brillouin zone can be twofold degenerate. 

\subsection*{Peculiarity of the cosine potential} In general, band crossings at the center of the BZ are allowed for symmetric potentials since the avoided crossing theorem does not apply, but usually only some band crossings appear while other bands still avoid each other. The simple cosine potential is special in the sense that  
 all eigenvalues to $q=0$ are twofold degenerate except for the ground state $\lambda_{00}=0$. 
 
 This property is somewhat hard to see in the representation of the time-evolution operator in the Schr\"odinger representation in the Fourier basis $\langle \mu | \mathcal{L}_0 | \nu\rangle$. However, the property can be easily deduced, omitting the gauge transform in the first place, i.e. representing the dynamics in terms of the non-Hermitian matrix 
\begin{align}
\langle \mu | \Omega | \nu\rangle =-  \frac{4\pi^2 D}{L^2}  \left[ \mu^2  \delta_{\mu\nu}  + \frac{\mu}{2 } ( \delta_{\mu,\nu+1} - \delta_{\mu,\nu-1} ) \right].
\end{align}
This matrix displays the symmetry $ \langle -\mu | \Omega | - \nu \rangle = \langle \mu | \Omega | \nu \rangle^*$. The argument now follows the one of 
Appendix  A of Ref.~\cite{Rusch_2024}. The matrix $\langle \mu | \Omega | \nu\rangle$ displays a zero row for $\mu=0$ and splits into a part  with entries for $\mu>0, \nu \geq $ and an identical one for $\mu<0, \nu \leq 0$.  

The only entries preventing the matrix to split into blocks with positive/negative $\mu, \nu$ are the matrix elements $\langle \pm 1 | \Omega| 0 \rangle$. However, 
as $\langle l_{0} | = \langle 0 |$ is a left eigenvector to eigenvalue $0$, all eigenvectors $| r_{n0} \rangle$ to non-zero eigenvalues have a zero entry in their Fourier representation by orthogonality of eigenvectors $0 = \langle l_{00} | r_{n0} \rangle = \langle 0 | r_{n0} \rangle$. Therefore, the blocks with both $\mu,\nu$ positive do not communicate with the blocks with both indices negative. In particular, one can choose eigenvectors with $\langle \mu | r_n \rangle = 0$ for $\mu \gtrless 0$ or symmetric and antisymmetric eigenfunctions to the twofold degenerate eigenvalue $\lambda_{n0} > 0$.

\section{Appendix: ISF  and probability density} \label{app:ISF_simplification}

Starting with the general definition of the ISF, \cref{eq:ISF_integral}, and inserting the expression for the probability density, \cref{eq:probability_density_S}, we obtain
\begin{align} 
	F_{\mu \nu}(q,t) =& \int_0^{L} \! \diff x_0 \int_0^{NL} \! \diff x \, e^{ -i(q+Q_\mu)x } e^{i(q+Q_\nu)x_0} 
\sqrt{\frac{p^\text{eq}(x)}{p^\text{eq}(x_0)}} 
 \frac{1}{N}  \sum_{q^\prime \in \text{BZ}} \sum_{n}  e^{-\lambda_{nq^\prime}t}  e^{iq^\prime(x-x_0)} u_{nq^\prime}(x) u_{nq^\prime}(x_0)^*  p^{\text{eq}}(x_0) \nonumber \\
	=& \int_0^{L} \! \diff x_0 \int_0^{NL} \! \diff x \, e^{ -i(q+Q_\mu)x } e^{i(q+Q_\nu)x_0}  \frac{1}{N} 
\sum_{q^\prime \in \text{BZ}} \sum_{n} e^{-\lambda_{nq^\prime}t}  e^{iq^\prime(x-x_0)} u_{nq^\prime}(x) u_{00}(x)^* u_{nq^\prime}(x_0)^*  u_{00}(x_0) \nonumber \\
=& \sum_{n}  e^{-\lambda_{nq}t} 
\left[ \int_0^{L} \! \diff x \, e^{ -iQ_\mu x }  u_{nq}(x) u_{00}(x)^*  \right] 
\left[ \int_0^{L} \! \diff x_0 \, e^{-iQ_\nu x_0} u_{nq}(x_0)  u_{00}(x_0)^*  \right]^* .
\nonumber \\
\end{align}
Here   in the second equality we used that the Bloch function to wave vector zero at the lowest band is related to the equilibrium density
$ u_{00}(x) = \sqrt{p^{\text{eq}}(x)}$. Furthermore, we observed that  the integral vanishes for $q\neq q^\prime$ due to the periodicity of the Bloch functions. 

We can make further progress by using the Fourier modes as basis functions, \cref{eq:Fourierexpansion_p}.
By expanding the Bloch functions we obtain 
\begin{align}
&  \int_0^{L} \! \diff x \, e^{-i Q_\mu x_0 } u_{nq}(x) u_{00}(x)^* = 
\int_0^{L} \! \diff x \, e^{-i Q_\mu x } u_{nq}(x_0)  \left[ 
 \sum_{\sigma\in \mathbb{Z}}  \frac{e^{i Q_\sigma x}}{\sqrt{L}} \langle \sigma | u_{00} \rangle  \right]^* \nonumber \\
 &= \sum_{\sigma\in \mathbb{Z}}  \langle u_{00} | \sigma  \rangle\int \frac{\diff x}{\sqrt{L}} e^{-i Q_{\mu+\sigma} x} u_{nq}(x)   = 
 \sum_{\sigma\in \mathbb{Z}}  \langle u_{00} | \sigma  \rangle \langle \mu + \sigma | u_{nq} \rangle .
		\end{align}  
Collecting terms, we find the expression for the ISF 
		\begin{align} \label{eq:ISF_fouriermodes}
			F_{\mu \nu}(q,t) 
		&= \sum_{n }  e^{-\lambda_{nq}t} \sum_{\sigma, \tau \in \mathbb{Z}} \langle u_{00} | \sigma \rangle \langle \mu+ \sigma | u_{nq} \rangle   \langle u_{nq} |\nu + \tau \rangle \langle \tau | u_{00} \rangle .
		\end{align}
This relation is \cref{eq:ISF_braket} in the main text.

We also derive how the probability density can be obtained from the intermediate scattering function
\begin{align}
&\frac{1}{N L^2 } \sum_{\mu\nu\in \mathbb{Z}} \sum_{q\in \text{BZ} }  F_{\mu\nu}(q,t ) e^{i (q+ Q_\mu) x} e^{-i (q+Q_\nu) x_0} = 
\frac{1}{N L^2 } \sum_{\mu\nu\in \mathbb{Z}} \sum_{q\in \text{BZ} } \langle  e^{- i (q+ Q_\mu) [ x(t)- x]} e^{i (q+Q_\nu) [x(0)-x_0]} \rangle \nonumber  \\
&=\frac{1}{N L^2 } \sum_{\mu\nu\in \mathbb{Z}} \sum_{q\in \text{BZ} } \sum_{q^\prime \in \text{BZ}}  \langle  e^{- i (q+ Q_\mu) [ x(t)- x]} e^{i (q^\prime+Q_\nu) [x(0)-x_0]} \rangle = 
N \int \frac{\diff q}{2\pi} \int \frac{\diff q^\prime}{2\pi}   \langle  e^{- i q [ x(t)- x]} e^{i q^\prime [x(0)-x_0]} \rangle \nonumber \\
&= N \langle \delta[x- x(t) ] \delta[x_0 - x(0) ] \rangle = p^{\text{eq}}(x_0) \mathbb{P}(x,t | x_0) .
\end{align}
This is \cref{eq:prob_by_ISF}  of the main text. 

\section{Harmonic approximation} \label{app:HA}
The harmonic approximation of the Langevin equation, \cref{eq:Langevin}, is given by
\begin{equation}
\frac{d}{dt} \bar x(t) = - D u Q_1^2  \bar x(t) + \eta(t),
\end{equation}
where the first term on the right-hand side is the restoring force, and the second term is the random force of the Brownian motion. The equation is expressed by the shifted position, $\bar{x} \coloneq x - L/2$, so that the potential minimum is at the center. 
The Smoluchowski equation to solve for the propagator $\mathbb{P} \coloneq \mathbb{P}(\bar x t \mid \bar x_0)$ is given by 
\begin{equation}
\partial_t \mathbb{P} = \frac{\partial}{\partial \bar x} ( D u Q_1^2  \bar x \mathbb{P} ) + D \frac{\partial^2}{\partial \bar x^2} \mathbb{P}
\end{equation}
with the harmonic-well relaxation time  $\tau = 1 /D u Q_1^2 $. The solution is known as the Ornstein-Uhlenbeck process \cite{risken1996fokker} and we find the  propagator 
\begin{align}
    \mathbb{P}(\bar x t \mid \bar x_0)=\frac{1}{\sqrt{2 \pi V(t)}} \exp \left[-\frac{\left(\bar x-\bar x_0 e^{-t / \tau}\right)^2}{2 V(t)} \right],
\end{align}
with $V(t)=D \tau\left[1-\exp{(-2 t / \tau)}\right]$,
and the stationary solution as the long-time limit of the propagator
\begin{align}
    p^\text{eq}(\bar x_0)=\frac{1}{\sqrt{2 \pi D \tau}} e^{-\bar x_0^2/2 D \tau}.
\end{align}

We can readily compute the generalized ISF by using the definition of the main text, \cref{eq:ISF_integral}, and by extending the integrals to infinity
\begin{align} \label{eq:ISF_integral1}
		F_{\mu\nu}(q,t) =& \int_\infty^{\infty} \! \diff \bar x \int_\infty^{\infty} \! \diff \bar x_0 \, e^{-i(q+Q_\mu) (x+L/2 )} e^{i(q+Q_\nu) (x_0+L/2)}  \mathbb{P}(\bar x, t | \bar x_0 )  p^{\text{eq}}(\bar x_0) .
\end{align}
Solving the integrals then yields the ISF
\begin{align}\label{eq:ISF_HA} 
   F_{\mu \nu}(q,t)=&  \exp \left[-\frac{D \tau }{2}  \left[(q+Q_\mu)^2-2 (q+Q_\mu) (q+Q_\nu) e^{-\frac{t}{\tau }} +(q+Q_\nu)^2\right]-\frac{i L}{2}  (Q_\mu-Q_\nu)\right].
\end{align}

We further calculate the ratio 
\begin{align}\label{eq:ISF_HA_ratio} 
\frac{F_{\mu \nu}(q,t \to \infty)}{F_{\mu \nu}(q,t)}=&  \exp {\left[ -\frac{k_B T }{U_1 Q_1^2} (q+Q_\mu) (q+Q_\nu) \right]}. 
\end{align}

We also readily find the MSD
\begin{align}\label{eq:MSD_HA} 
    \langle \Delta x(t)^2 \rangle &=\int_\infty^{\infty} \! \diff \bar x \int_\infty^{\infty} \! \diff \bar x_0 \,(x-x_0)^2 \times \mathbb{P}(\bar x, t | \bar x_0 )   p^{\text{eq}}(\bar x_0) = 2 D \tau\left(1-e^{- t / \tau}\right),
\end{align}
and time-dependent diffusion coefficient  
\begin{align}
    D(t)    &=  De^{- t / \tau}.
\end{align}


\end{widetext}

\providecommand*{\mcitethebibliography}{\thebibliography}
\csname @ifundefined\endcsname{endmcitethebibliography}
{\let\endmcitethebibliography\endthebibliography}{}

\end{document}